%% file: europium.tex
\documentclass[10pt,twocolumn,floatfix,aps,pra,superscriptaddress,]{revtex4-1} 
\usepackage{amsmath,amssymb,amsthm,mathrsfs,amsfonts,dsfont,amstext} 
\usepackage{textcomp,pbox}
\usepackage{bm}
\usepackage{dcolumn,booktabs,url} 
\usepackage[scaled]{helvet}
\usepackage{sansmath,gensymb}
\usepackage{tikz,graphicx,transparent,color}
\usepackage{multirow}
\usepackage{physics}

\usepackage[colorlinks=true]{hyperref}

\graphicspath{{./figures/}}

\newcommand{\yso}{Y$_2$SiO$_5$}
\newcommand{\eu}[0]{Eu$^{3+}$:Y$_2$SiO$_5$}
\newcommand{\euiso}[0]{$^{151}$Eu$^{3+}$:Y$_2$SiO$_5$}
\newcommand{\transition}{$^7$F$_{0} \longleftrightarrow ^5$D$_{0}$ }
\newcommand{\figref}[1]{\figurename{~\ref{#1}}}
\newcommand{\tabref}[1]{\tablename{~\ref{#1}}}
\newcommand{\mathbbm}[1]{\text{\usefont{U}{bbm}{m}{n}#1}}

\begin{document}

\newcommand{\TitleName}{Characterization of the hyperfine interaction of the excited  $^5$D$_0$ state of \eu{}}

\newcommand{\AffGeneve}{Groupe de Physique Appliqu\'ee, Universit\'e de Gen\`eve, CH-1211 Gen\`eve, Switzerland}
\newcommand{\AffParis}{PSL Research University, Chimie ParisTech, CNRS, Institut de Recherche de Chimie Paris, 75005 Paris, France}

\title{\TitleName}

 \author{Emmanuel Zambrini Cruzeiro}
 \affiliation{\AffGeneve{}}
 \author{Jean Etesse}
  \affiliation{\AffGeneve{}}
 \author{Alexey Tiranov}
  \affiliation{\AffGeneve{}}
  \author{Pierre--Antoine~Bourdel} 
   \affiliation{\AffGeneve{}}
 \author{Florian Fr\"owis} \affiliation{\AffGeneve{}}
 \author{Philippe Goldner} 
  \affiliation{\AffParis{}}
 \author{Nicolas Gisin}
  \affiliation{\AffGeneve{}}
 \author{Mikael Afzelius} \email[Email to: ]{mikael.afzelius@unige.ch}
  \affiliation{\AffGeneve{}}

\date{\today}

\begin{abstract}
We characterize the europium (Eu$^{3+}$) hyperfine interaction of the excited state ($^5$D$_0$) and determine its effective spin Hamiltonian parameters for the Zeeman and quadrupole tensors. An optical free induction decay method is used to measure all hyperfine splittings under weak external magnetic field (up to 10~mT) for various field orientations. On the basis of the determined Hamiltonian we discuss the possibility to predict optical transition probabilities between hyperfine levels for the \transition{} transition. The obtained results provide necessary information to realize an optical quantum memory scheme which utilizes long spin coherence properties of \euiso{} material under external magnetic fields.
\end{abstract}

\maketitle 

\section{INTRODUCTION}

Rare-earth-ion-doped crystals (REIC) have been actively studied during the last decade as promising solid-state materials for quantum information processing.  In the field of quantum communication these compounds have been used as optical quantum memories: devices capable to store and release  quantum states of light~\cite{Lvovsky2009,Tittel2010b,Bussieres2013,RiedmattenAfzeliusChapter2015a}. In this context different quantum memory protocols were utilized to demonstrate high-efficiency \cite{Hedges2010,Sabooni2013,Jobez2014}, long storage time \cite{Lovric2013,Heinze2013,Laplane2016a}, efficient temporal \cite{Usmani2010} and frequency multiplexing \cite{Saglamyurek2015a}, multiple-photon storage and entanglement storage \cite{Clausen2011,Saglamyurek2011} in various types of REICs.

Europium-doped yttrium orthosilicate \eu{} is one of the most attractive solid-state systems to realize optical quantum memory for quantum repeater application. This is due to the long optical coherence times of a few milliseconds \cite{Yano1991,Equall1994,Koenz2003} which together with excellent spin coherence properties of tens of milliseconds lifetime \cite{Alexander2007} offer the possibility to realize spin-wave storage of photonic states \cite{Afzelius2010}. Storage times up to a few milliseconds have been demonstrated using different quantum memory schemes at zero magnetic field \cite{Jobez2015,Laplane2016a,Laplane2017}.

Recently, the extension of the spin coherence lifetime in \eu{} up to one minute has been demonstrated using the zero-first-order Zeeman shift (ZEFOZ) condition at high magnetic fields \cite{Zhong2015}. The long coherence time is due to the decoupling of the hyperfine transition from magnetic-field fluctuations from the host spin flips \cite{Fraval2004,Fraval2005}. Further application of the dynamical decoupling technique using trains of rf-pulses resulted in extended hyperfine coherences up to 6 hours \cite{Arcangeli2014,Zhong2015}. This clearly demonstrates the potential of \eu{} crystals to realize long-duration quantum light matter interface applicable for quantum communication.

Further use of the ZEFOZ transition for storing optical excitations requires knowledge of optical properties for this material under an applied magnetic field. This is important in order to find a proper energy level path where single photons can be efficiently transferred to spin-wave excitations. The excited state spin Hamiltonians have been previously characterized for other non-Kramers crystals \cite{Lovric2011,Lovric2012}. This information allowed to predict optical transition probabilities  between the ground and excited hyperfine levels. However, the magnetic properties of the $^5$D$_0$ excited state of \eu{} crystal have not been fully characterized so far.

In this work, we investigate the hyperfine properties of the excited state $^5$D$_0$ of \euiso{} by fully reconstructing its effective spin Hamiltonian.  To this end we use an optical free induction decay (FID) method on the optical \transition{} transition \cite{Koenz2003}, which allows us to measure all hyperfine splittings under weak external magnetic fields (up to 10~mT) applied in various directions. With this approach, all hyperfine splittings can be measured for both the ground and excited states at the same time, which is an  efficient method to precisely characterize the relative orientation of the two spin Hamiltonians (for ground and excited states). This is crucial in order to predict optical branching ratios for various optical pumping tasks, like quantum memory applications. Using both Hamiltonians, we are able to find parameters that result in an good agreement between calculated and experimental optical transition probabilities the optical transition probabilities for different hyperfine levels of the optical \transition{} transition as a function of the external magnetic field.  

The work is organized as follows. In Section~\ref{sec:background}, we present the effective Hamiltonian describing the magnetic properties of hyperfine levels in \euiso{}. In Section~\ref{sec:exp}, we present the measurement method and the experimental details. Section~\ref{sec:results} shows the main results: the measurement of ground state and excited state hyperfine splittings as a function of the external magnetic field's angle and the prediction of the transition probabilities using the fitted parameters. We finally discuss the implications of our findings and give an outlook in Section~\ref{sec:conclusion}. 

\section{Hyperfine interaction for REICs}
\label{sec:background}
\subsection{Spin Hamiltonian}

The hyperfine interaction of rare-earth  centers is usually described using a Hamiltonian of the form~\cite{MacfarlaneShelby1987}
\begin{equation}\label{eq:H0}
\mathcal{H_0}=\left[\mathcal{H}_{\mathrm{free}}+\mathcal{H}_{\mathrm{cf}}\right] +  \left[\mathcal{H}_{\mathrm{hyp}} + \mathcal{H}_{\mathrm{Q}}  + \mathcal{H}_{\mathrm{Z}}  + \mathcal{H}_{\mathrm{z}} \right],
\end{equation}
where the first two terms describe the free ion and the crystal-field (cf), which together characterize the electronic coupling and determine the optical transitions. All other terms describe the hyperfine coupling, the nuclear quadrupole coupling, and the electronic and nuclear Zeeman Hamiltonians, respectively.

\begin{figure}[h!]
\includegraphics[width=.9\linewidth] {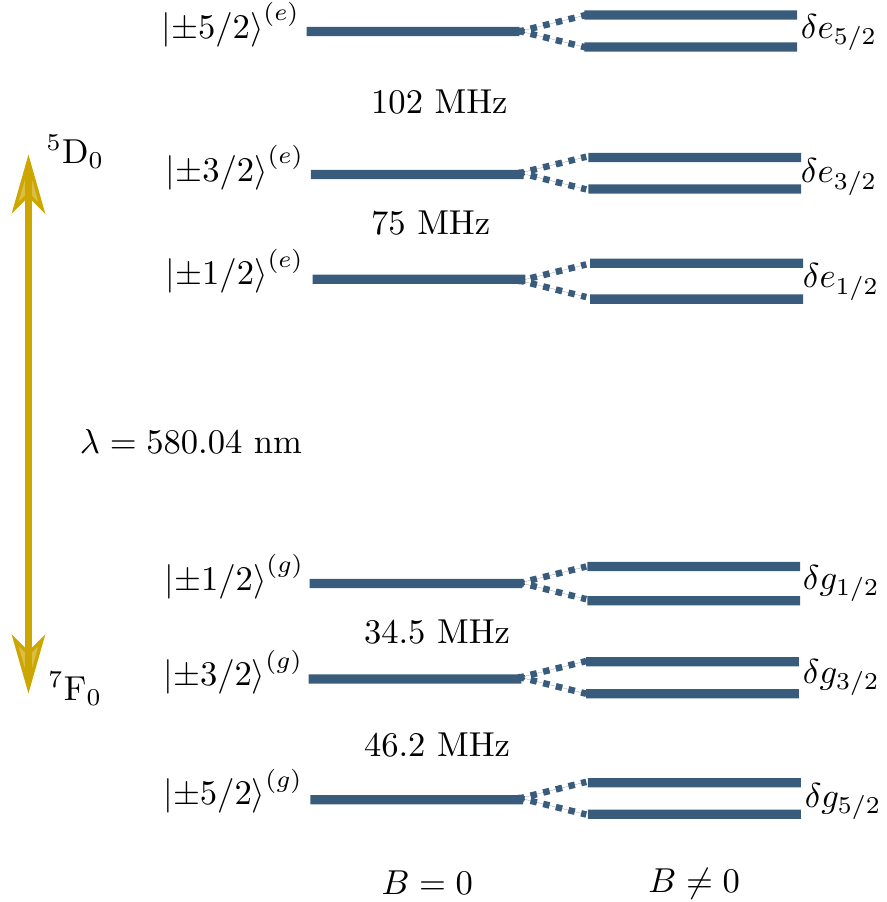}
\caption[]{\label{fig:levels} (color online)
The energy level structure of \euiso, without (left) and with (right) an external magnetic field $B$. The inhomogeneous broadening of the optical transition \transition{} resonant at 580~nm contains hyperfine $m_I = - 5/2 \dots +5/2$ sub-levels both for the ground $^7$F$_0$ and the excited $^5$D$_0$ states. The ground state hyperfine splittings were characterized in \cite{Longdell2006}.}
\end{figure}

In the present work we consider the optical transition of $^{151}$Eu$^{3+}$ between the ground $^7$F$_0$ (denoted as $\ket{g}$) and the excited state $^5$D$_0$ (denoted as $\ket{e}$), which for \yso{} material takes place at 580.04~nm wavelength (in vacuum, optical site I \cite{Yano1991}). The energy level structure is displayed in \figref{fig:levels}.

Due to the singlet states ($J=0$) connected by the optical \transition{} transition for Eu$^{3+}$ ion and the even number of electrons the net orbital angular momentum and the electron spin are quenched \cite{MacfarlaneShelby1987}. 
This allows to efficiently represent the second group of terms in Eq.~\eqref{eq:H0} as a perturbation for the electronic levels. Due to the quenching, the hyperfine coupling and electronic Zeeman interactions are not present at the first order, which at zero magnetic field leads to the same order of magnitude for all the terms inside the second brackets of~Eq.~\eqref{eq:H0}.

Representing these terms as a second order perturbations for the first group allows us to consider only the effective nuclear spin Hamiltonian \cite{Teplov1968, Longdell2002}
\begin{equation}
\mathcal{H}_{\mathrm{}}=\hat{I}\cdot \mathbf{Q}\cdot\hat{I} + \vec{B}\cdot \mathbf{M}\cdot \hat{I}+(\vec{B}\cdot\mathbf{Z}\cdot\vec{B})\mathbbm{1}.
\label{eq:Heff}
\end{equation}
In this expression, the first term corresponds to the quadrupole interaction and is responsible for a partial lifting of the nuclear-spin states degeneracy in both the ground and the excited states for the $I=5/2$ nuclear spin of europium (see  \figref{fig:levels}, left). In general, this term includes pure quadrupolar and pseudoquadrupolar contributions \cite{Teplov1968}.  The second term describes the Zeeman interaction and results in non-degenerate hyperfine levels in the presence of a magnetic field (see \figref{fig:levels}, right). The third term is the quadratic Zeeman interaction, which we neglect since it does not contribute to the admixtures of the eigenstates. The labels used for the hyperfine levels in \figref{fig:levels} are only approximate, since $m_I$ is not a good quantum number. 

As the energy splittings due to $\mathcal{H}_{\mathrm{}}$ are very small compared to the optical transition, this term can be seen as a perturbation of the whole Hamiltonian. Two hyperfine Hamiltonians can be defined: one for the ground state $\mathcal{H}_{\mathrm{}}^{(g)}$ and one for the excited state $\mathcal{H}_{\mathrm{}}^{(e)}$. The hyperfine ground state Hamiltonian has already been determined in a previous work \cite{Longdell2006}. We are thus interested in the present work in characterizing the Hamiltonian of the excited state and its orientation with respect to the ground state Hamiltonian. This is done by determining experimentally $\mathbf{Q}^{(e)}$ and $\mathbf{M}^{(e)}$, that is, the quadrupole and Zeeman tensors of the excited state hyperfine Hamiltonian. 

\subsection{Symmetry considerations in \yso{}}

In the present work we study only one of the stable europium isotope, particularly $^{151}$Eu. While two isotopes $^{151}$Eu and $^{153}$Eu appear in approximately equal concentrations, their magnetic properties are slightly different. The larger electric quadrupole moment of $^{153}$Eu usually results in larger zero field splittings, while nuclear gyromagnetic ratio is usually stronger for the $^{151}$Eu isotope~\cite{Erickson1981, Liu2005a}. For quantum information applications the $^{153}$Eu isotope can offer a larger optical bandwidth and potentially longer coherence times, however the magnetic field intensities required to find ZEFOZ transitions are larger with this isotope, due to the stronger electric quadrupole moment~\cite{Zhong2015}.

\yso{} is a monoclinic biaxial crystal of the $C_{2h}^6$ space group. When Eu$^{3+}$ ions substitute yttrium Y$^{3+}$ ions they can occupy two different crystallographic sites. Here we study the crystallographic site which offers a higher absorption coefficient and a longer optical coherence time (site I)~\cite{Yano1991}. For this site, europium ion can also occupy two magnetically inequivalent subsites, and the Hamiltonians of these two subsites are related by a $\pi$-rotation around the $C_2$ symmetry axis of the crystal. This means that two quadrupole tensors $\mathbf{Q}^{(e)}_1$ and $\mathbf{Q}^{(e)}_2$ and two Zeeman tensors $\mathbf{M}^{(e)}_1$ and $\mathbf{M}^{(e)}_2$ must be defined, one per magnetic subsite. 
Note that the two magnetic subsites become equivalent when an external magnetic field is applied along the crystal symmetry $C_2$ axis  or in the plane perpendicular to it.  The crystal was cut along the polarization extinction axes $D_1, D_2$ and $b$ \cite{Li1992}, where $b$ coincides with the crystallographic $C_2$ symmetry  axis.

To summarize, in this work we determine the two tensors $\mathbf{Q}^{(e)}_1$ and $\mathbf{M}^{(e)}_1$ in the ($D_1$, $D_2$, $b$) basis by measuring the splittings of the hyperfine excited state due to the presence of a magnetic field. The two tensors $\mathbf{Q}^{(e)}_2$ and $\mathbf{M}^{(e)}_2$ are then deduced by a $\pi$-rotation around the $C_2$ axis (see Appendix~\ref{app:tensors} for more details). Since the point symmetry at the site of Eu$^{3+}$ in \yso{} crystal is $C_1$, the tensor axes for each interaction type can be arbitrarily oriented with respect to each other for a given electronic state, and additionally have different relative orientations in the ground and excited states. This makes the characterisation of their relative orientations in different electronic states a complicated problem.

\section{EXPERIMENTAL METHODS}
\label{sec:exp}

Several experimental methods can be used to measure the ground and excited state splittings. The most common techniques combine optical and radio-frequency (rf) fields, such as Raman Heterodyne Scattering (RHS)~\cite{Longdell2006,Lovric2012,Ahlefeldt2013b}. This method requires an efficient coupling between rf-radiation and the spin transition under study. Due to the large quadrupole splittings and the weak Rabi frequencies for the excited state $^5$D$_0$, this method is technically demanding in terms of rf power and impedance matching. Preliminary RHS signals we recorded were weak and difficult to use for a quantitative analysis. This does not preclude the use of RHS for such a measurement, however we chose another approach to obtain the required experimental data.

To overcome these technical limitations, we use spectral hole burning (SHB). With SHB, one can measure simultaneously the ground and excited state splittings with a single absorption measurement and without using rf fields. A difficulty using SHB is the interpretation of the complicated SHB spectrum. To solve this problem we use a technique called class cleaning, which we now describe in detail.

\subsection{Class cleaning for SHB at the Zeeman level}
\label{subsubsec:shb}
 
\begin{figure}[h!]
\includegraphics[width=0.9\linewidth] {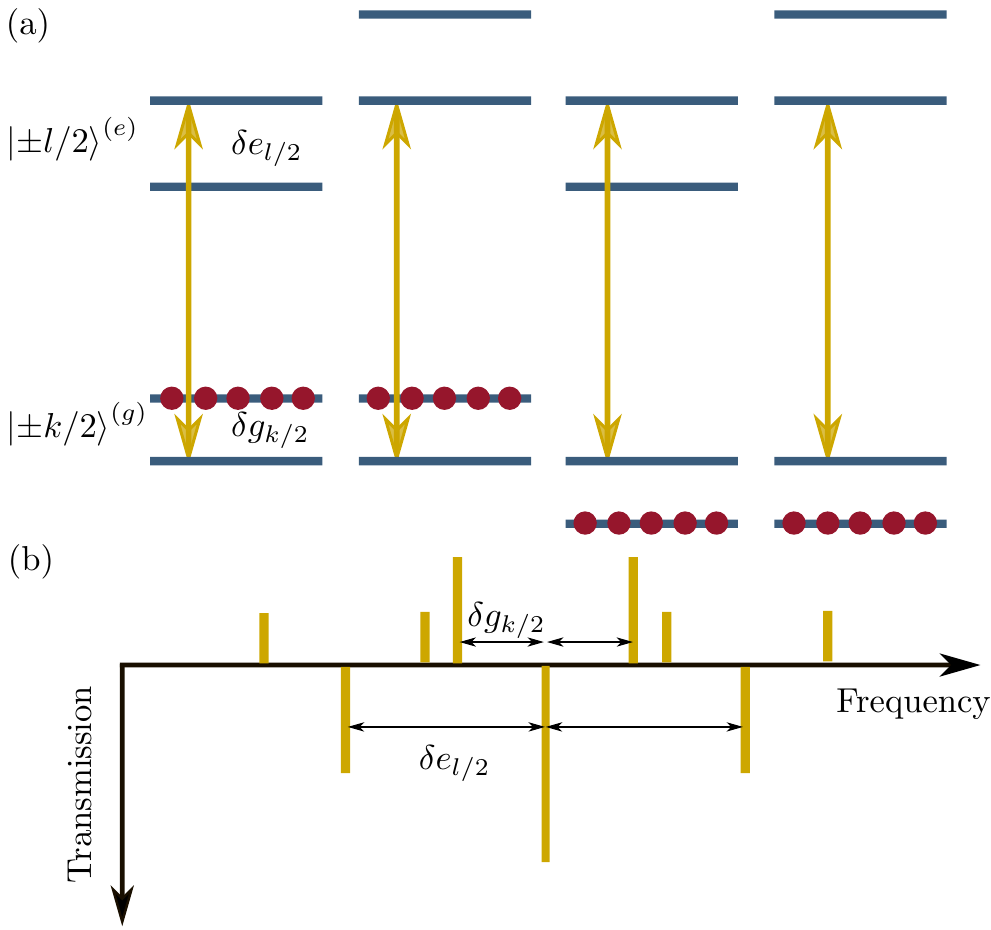}
\caption[]{\label{fig:classes} (color online)
(a) The class cleaning procedure at the quadrupole level (described in the main text) leaves 4 classes of atoms between the ground $|\pm\frac{k}{2}\rangle^{(g)}$ and excited $|\pm~\frac{l}{2}\rangle^{(e)}$  Zeeman doublets. Under an external magnetic field, each doublet splits with energy difference $\delta e$ and $\delta g$ for the ground and excited state respectively. (b) Using SHB technique one can redistribute the population to reveal the energetic structure with contribution from the 4 different classes.}
\end{figure}

The general idea of SHB is the following \cite{Macfarlane1987,Liu2006}: given that the inhomogeneous broadening of the \transition{} transition is large compared to the hyperfine splittings, sending a pump laser of fixed frequency on the ensemble for a much longer time than the radiative lifetime will cause the atoms to be redistributed among the hyperfine ground state levels. For a system with $N_g$ ground state levels and $N_e$ excited states, there will be a total of $N_g\times N_e$ resonant transitions, corresponding to different classes of atoms. For instance, \figref{fig:classes}(a) shows the four classes of resonant atoms in the case $N_g=N_e=2$. The pumping process eventually leads to a spectral pattern of holes and antiholes in the absorption profile, shown in \figref{fig:classes}(b).

We could try to directly use this technique to probe the different splittings we want to measure, but the spectral pattern for $I=5/2$ would be composed of 31 holes and 930 anti-holes originating from the 36 classes of atoms for each magnetically inequivalent site. Retrieving the excited and ground state splittings would be a challenging task in this case.

Instead of using all the 36 classes of atoms, we perform a class cleaning of the atoms at the quadrupole level \cite{Nilsson2004}. This means that by using an appropriate sequence presented in detail in \cite{Lauritzen2012,Laplane2016a} we  address a single transition of the kind $\ket{\pm k/2}^{(g)}\longleftrightarrow \ket{\pm l/2}^{(e)}$, with $(k,l)\in \{1,3,5\}$. Since the class cleaning is only done at the quadrupole level, when sending light on a $\ket{\pm k/2}^{(g)}\longleftrightarrow \ket{\pm l/2}^{(e)}$ transition, we simultaneously address atoms on the four transitions associated with this system. This is due to the fact that all the Zeeman splittings ($\sim 200\ \mathrm{kHz}$) are much smaller than the bandwidth of the class cleaning procedure ($5\ \mathrm{MHz}$). Hence, we are left with 4 classes of atoms on the Zeeman structure instead of 36, as depicted in \figref{fig:classes}(a). Once this class cleaning procedure has been performed, burning a hole leads to the appearance of 3 holes and 6 anti-holes, the positions of which directly give the excited $\delta e$ and the ground state splittings $\delta g$ \cite{Hastings-Simon2008a} (see \figref{fig:classes}(b)). For typical Zeeman splittings lower than 400~kHz the challenge of measuring holeburning spectra is twofold: first the burning laser should have a narrower linewidth than the energy splittings, and second the readout of the structure should be very precise in order to resolve it. We will see in the next section how we solve these issues in the present case.

\subsection{SHB spectrum measurement: heterodyne measurement of the FID}
\label{subsubsec:fid}

\begin{figure*}
\includegraphics[width=1\linewidth] {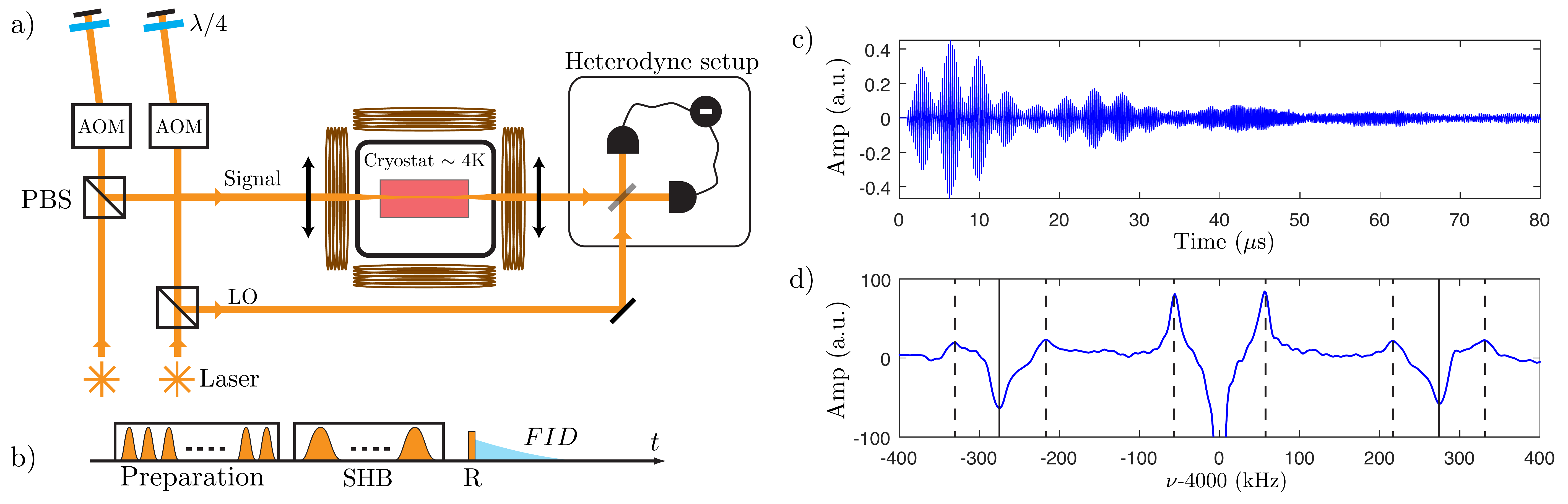}
\label{fig:exp_spec}
\caption[]{\label{fig:setup} (color online)
(a) Experimental setup. The laser is split into two beams and sent to two different paths: the signal, which prepares and probes the crystal and the local oscillator (LO). The amplitude and frequencies of these beams are adjusted using acousto-optic modulators (AOM), which are used in double-pass configuration. PBS stands for polarizing beamsplitter. Coils around the cryostat provide the external magnetic field. (b) Experimental sequence, consisting of a preparation step, a spectral hole burning (SHB) step and the readout (R) step. The FID is measured with an oscilloscope right after the end of the readout pulse. (c) Oscilloscope trace for 10~mT magnetic field applied along $\vec{B} \parallel D_1$ : interference of the FID with the 4~MHz detuned local oscillator. The beatings reveal a complex absorption structure. (d) Imaginary part of the Fourier transform of the temporal trace presented in (c), which is proportional to the absorption of the spectral structure. Antiholes are indicated by a dashed vertical line, the solid vertical lines correspond to sideholes.}
\end{figure*}

As explained previously, our goal is to measure SHB spectra, like the one shown in \figref{fig:classes}(b), and extract the excited state splittings as a function of the direction of the magnetic field.
A first simple idea is to use a readout pulse, whose frequency is chirped over time. The limitation with this solution is that the resolution of the measurement is strongly linked with the chirp rate: as the structure that we want to measure is only a few kilohertz wide, the chirp rate should be very slow. This tends to work with very weak readout amplitudes to avoid hole burning due to the readout pulse, implying measurements with low signal-to-noise ratios.

Instead of a frequency-resolved absorption measurement, we  perform a temporal measurement of a signal emitted by the spectral structure we want to measure. In other terms, we excite the spectral structure with a short readout pulse, which will create an optical coherence on the atoms. These atoms will then emit light after the end of the readout pulse: This is the free induction decay (FID) \cite{Shelby1983,deSeze2005}. As a temporal counterpart of the direct spectral absorption measurement, the absorption spectrum is simply the imaginary part of the Fourier transform of the measured FID. This requires that the spectrum of the readout pulse should be large compared to the probed spectral structure.

To measure the FID, we use an interferometric technique called balanced heterodyne detection: we mix the FID field with a 4~MHz-detuned optical local oscillator (LO) on a 50:50 beamsplitter, and measure the difference in photocurrent of two photodiodes placed in its two outputs. The advantage of this method is that the measurement is only limited by the shot-noise of the readout pulse.

\subsection{Experimental setup}

In~\figref{fig:setup}(a) we show the experimental setup. Our laser source is a cavity-stabilized source with a sub-kHz linewidth, which emits 2~W of light at 580.04~nm. We use 40~mW for this experiment and split the power into two different beams. The first one, the signal beam, is used to prepare and excite the crystal sample. 
  The second beam is used as the local oscillator for the heterodyne detection. In order to modulate the frequencies and the amplitudes of both the signal and the LO for the implementation of the sequence, acousto-optical modulators (AOMs) in a double pass configuration are used. The AOMs are driven by an analog generator card that performs both amplitude and phase modulation. The signal beam is then recombined on a 50:50 beamsplitter with the local oscillator for the heterodyne measurement, performed by a balanced photodiode detector.

For our study we use a 1~cm long isotopically pure \euiso{} crystal with a doping concentration of 1000~ppm. We chose this particular host crystal for its low nuclear spin density, which leads to long optical and hyperfine coherence times \cite{Yano1991,Equall1994,Koenz2003}. The crystal was grown by the Czochralski method. For more details regarding the crystal and its growth, see Ref. \cite{Ferrier2016}.

To minimize the effect of decoherence processes, the crystal is cooled to 3~K in a commercial closed-cycle cooler from Cryomech, with a custom-made vibration-damping mount. In order to apply the magnetic field necessary to lift the Zeeman degeneracy, we use  three pairs of copper coils close to a Helmholtz configuration. The magnetic field is limited to $B_x=B_y=10\ \mathrm{mT}$ in the $X$ and $Y$ directions and to $B_z=5\ \mathrm{mT}$ in the $Z$ direction, due to heating through the Joule effect. The axes of the coils $X$, $Y$ and $Z$ define the lab frame in which the spin Hamiltonian is defined. The crystal axes $D_1$, $D_2$ and $b$ are oriented closely to the $Y$, $X$, and $Z$ axes of the coils, respectively.  Further possible misalignment is included in the fitting procedure discussed later.  

Each $\ket{\pm k/2}^{(g)} \longleftrightarrow \ket{\pm l/2}^{(e)}$ transition that is probed requires a specific preparation procedure: As we want the FID signal to be the strongest possible, we additionally polarize all the spins in the selected class to the $|\pm k/2\rangle^{(g)}$ state by optical pumping. These are simply variants of the basic class cleaning procedure discussed in Ref.~\cite{Laplane2016a}.

Figure~\ref{fig:setup}(b) shows the sequence that is used for the experiment. First the direction and amplitude of the magnetic field are set using three independent current sources. Then the atomic preparation occurs, which consists in the class cleaning procedure (see Section~\ref{subsubsec:shb}) and the pumping procedure previously mentioned. The preparation of the atoms is performed over an optical bandwidth of 5~MHz. Then, we perform SHB on the ensemble by sending a series of identical and spectrally narrow pulses. This sequence results in  burning  a structure of the type presented in \figref{fig:classes}(b), where the holes and anti-holes have a typical width of the order of 10~kHz. We believe that this width is currently limited by the residual vibrations of the crystal during the SHB procedure. Finally, a single 1.5~$\mu$s long square pulse is sent as the readout pulse. The beginning of the FID measurement is triggered right after the end of this pulse. The LO is continuously sent to the heterodyne detection, with a detuning of 4~ MHz with respect to the readout pulse.

\section{Experimental results}
\label{sec:results}

\subsection{Obtaining the absorption spectra}

Figure \ref{fig:setup}(c) shows a typical trace recorded by the oscilloscope: The FID is beating with the LO at 4~MHz, and the slow modulations reveal the existence of a structure in the spectral domain. Nevertheless, if we consider directly the imaginary part of the Fourier transform of the measured signal, we do not recover the expected absorption spectrum: In close analogy to NMR \cite{Ernst1969}, we need to apply a linear phase correction to our data. The origin of this phase correction is twofold: first, for each FID measurement the relative phase of the LO is random. A constant phase should then be added for each measurement. Secondly, the measurement does not start right at the beginning of the FID emission. This shift in time implies a linear correction in frequency. Once these corrections have been applied, we obtain the absorption profile shown in \figref{fig:setup}(d), which is of the same form as the one schematically presented in \figref{fig:classes}(b). 

\subsection{Scanning the magnetic field}

In order to reconstruct the two $\mathbf{Q}^{(e)}_1$ and $\mathbf{M}^{(e)}_1$ tensors, we have to know the splittings for several possible directions of the magnetic field. To scan the field homogeneously in space, we use the same method as the one presented in~\cite{Longdell2002}: we scan the magnetic field along a spiral parametrized by
\begin{equation}
\vec{B}_n=\left( \begin{array}{c} B_x\sqrt{1-t_n^2}\cos (6\pi t_n) \\ -B_y t_n \\ B_z\sqrt{1-t_n^2}\sin (6\pi t_n) \end{array} \right),
\end{equation}
where $t_n=-1+2\frac{n-1}{N-1}$, $n \in[\![1,N]\!]$. In our case, the scan of the space occurs along an ellipsoid, because $B_z\neq B_x=B_y$. Since the $D_1$--$D_2$ plane is roughly parallel to the $X-Y$ plane, if we scan around the $X$ or $Y$ axes we will cross the $D_1$-$D_2$ plane several times. Outside this plane we observe two different SHB spectra as shown in \figref{fig:classes}(b). Using this fact one can precisely identify the position of the $D_1$-$D_2$ plane from the spiral measurement. In all of the spiral measurements we present in the article, $N$ was chosen to be 200. 

In \figref{fig:spirals}, we show the SHB spectra for three transitions between the $^7$F$_0$ and $^5$D$_0$ manifolds, obtained with spiral scans. The hole positions were identified in these rotation patterns manually, by looking at the SHB spectrum for each orientation of the magnetic field along the spiral pattern individually. Whenever possible, the main antiholes would also be identified, however their amplitudes were generally smaller. 

\subsection{Fitting procedure}
\label{fit_section}

To find the Hamiltonian which explains the observed spectra, we parametrize the effective Hamiltonian (Eq.~\eqref{eq:Heff}). Since the diagonal elements of the quadrupolar tensor $\mathbf{Q}^{(e)}_1$ are known \cite{Mitsunaga1991}, we only fit the orientation of this tensor, using three Euler angles $\alpha_Q$, $\beta_Q$ and $\gamma_Q$ in the (X,Y,Z) lab frame. Then, the Zeeman part is described by six parameters. They  correspond to its three diagonal elements $g_1$, $g_2$ and $g_3$ and three angles $\alpha_M$, $\beta_M$ and $\gamma_M$ representing the orientation of the $\mathbf{M}^{(e)}_1$ tensor in the $(X,Y,Z)$ lab frame. These angles are not the same as for the $\mathbf{Q}^{(e)}_1$ tensor due to the low site symmetry in the crystal. Finally, two more parameters  are used to identify the orientation of the $C_2$ symmetry axis connecting  two magnetically inequivalent subsites: $\alpha_{C_2}$ and $\beta_{C_2}$ defined in spherical coordinates in $(X,Y,Z)$ lab frame.  A rotation of $\pi$ around this axis for both tensors is used to obtain the Hamiltonian for the second subsite containing $\mathbf{Q}^{(e)}_2$ and $\mathbf{M}^{(e)}_2$ tensors as explained in Sec.~\ref{sec:background}. The exact form of the Hamiltonian and details about the rotation transformations are given in the Appendix~\ref{app:tensors}. 

\begin{figure}
    \centering
    {\includegraphics[width=1\linewidth,trim=0.5cm 0.5cm 0.5cm 3.2cm,clip]{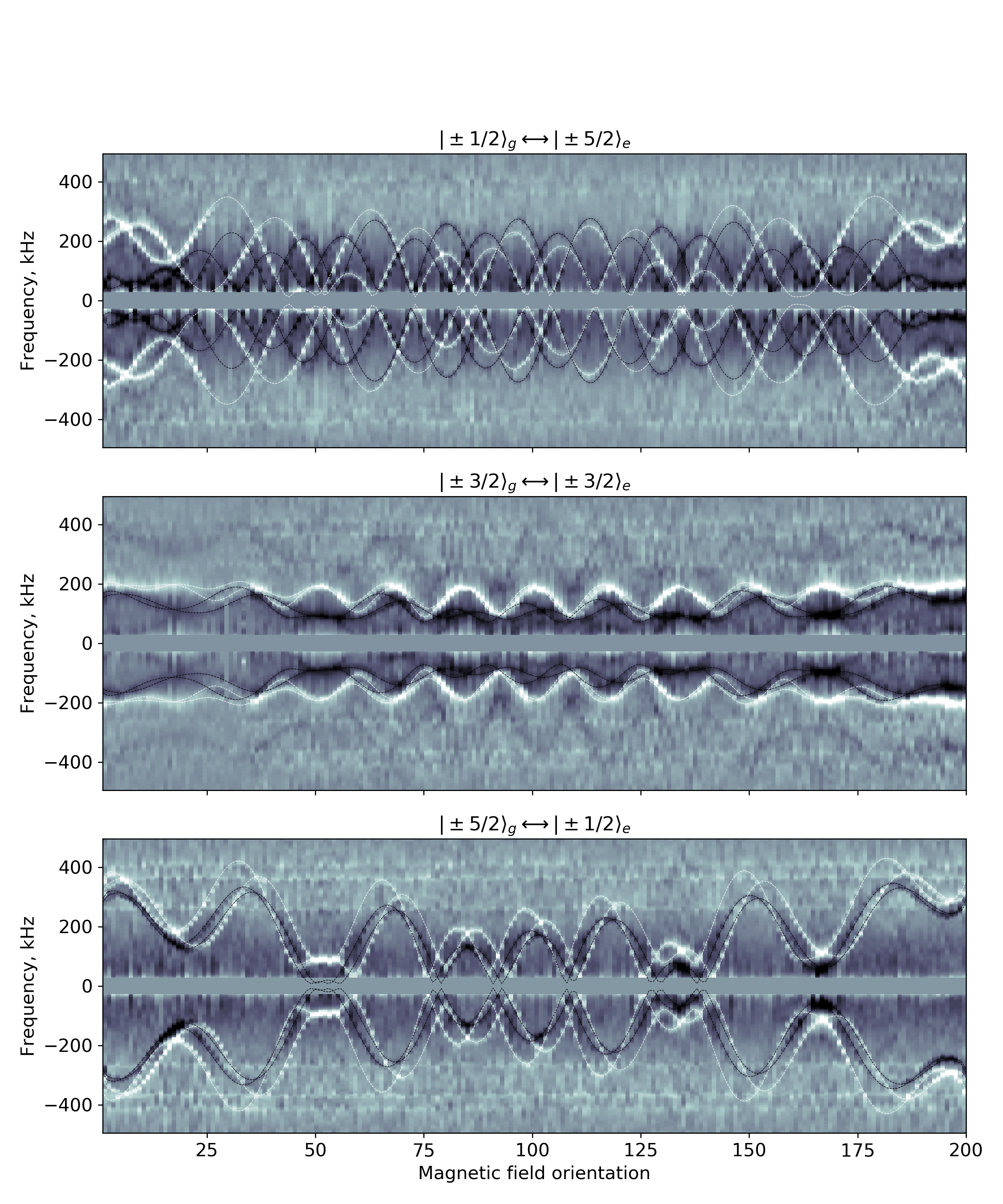}} 
    \caption{\label{fig:spirals} Experimental SHB spectra obtained using the spiral scan of the magnetic field, for transitions connecting different hyperfine levels of the $^7$F$_0$ ground state and the $^5$D$_0$ excited state of \euiso{}. Each vertical slice represents a hole burning spectrum obtained using the FID signal (an example is shown in \figref{fig:setup}(d)). White regions correspond to higher transmission (holes) while black regions represent increased absorption (antiholes). The positions of the side holes give directly the splittings of the excited state, while the positions of the strongest antiholes correspond to the ground state splitting (cf. \figref{fig:classes}). The energy splittings predicted by the fitted spin Hamiltonian are shown as white (holes) and black (anti-holes) lines. The strong central hole was removed to increase the contrast of the image. The colour axis is non-linear.}
\end{figure}

\begin{figure}
    \centering
    {\includegraphics[width=0.97\linewidth,trim=0.5cm 0.6cm 0.5cm 2.8cm,clip]{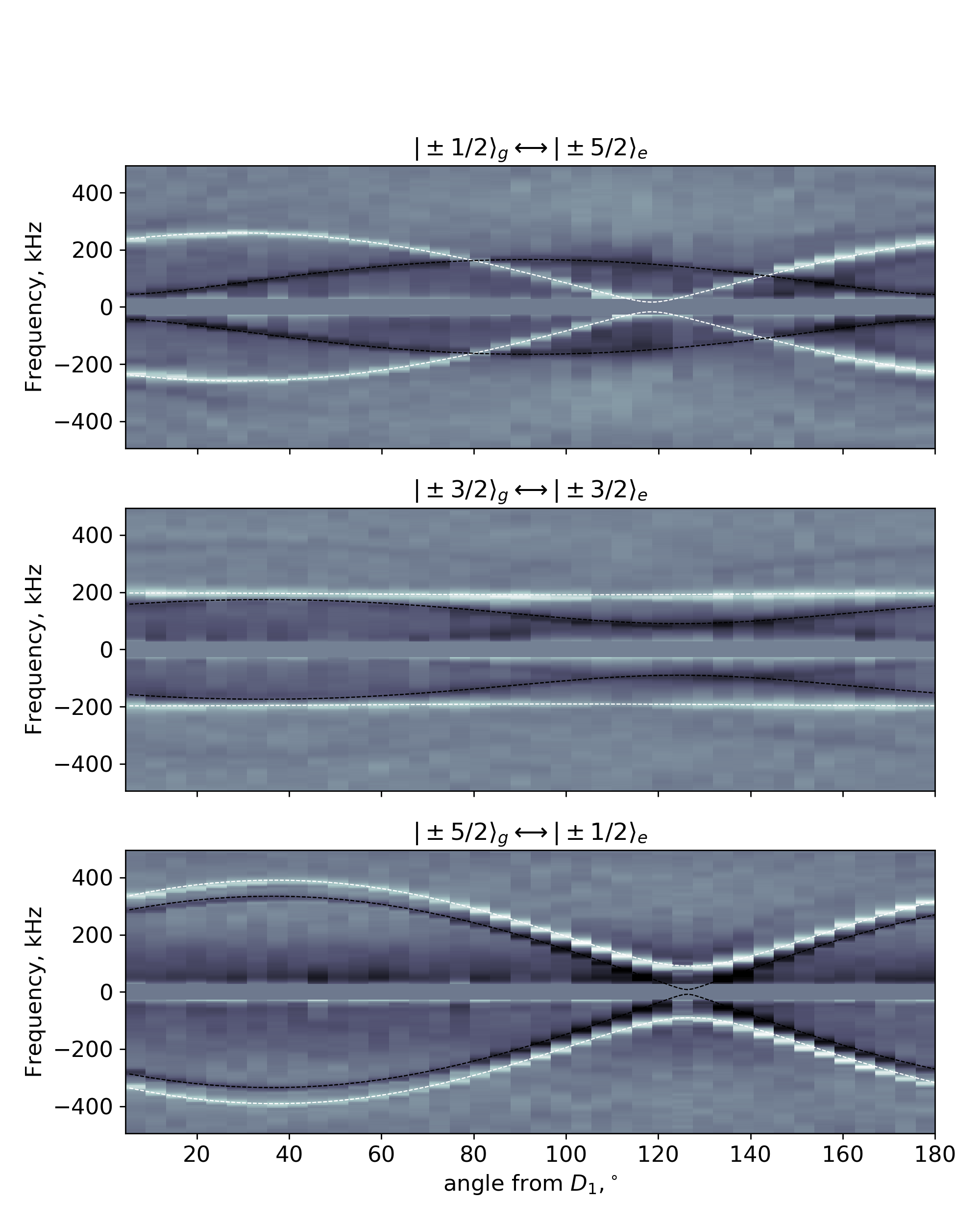}}
    \caption{\label{fig:D1D2b} Experimental SHB spectra obtained by scanning the magnetic field of 10~mT in the $D_1$-$D_2$ plane (perpendicular to the $C_2$ symmetry axis). In this plane, the magnetic subsites are degenerate, such that fewer holes and anti-holes are seen. The measured spectra are in good agreement with the energy splittings predicted by the fitted spin Hamiltonian, without any additional tuning of the parameters with respect to the fit shown in \figref{fig:spirals}. All other experimental details are identical to \figref{fig:spirals}.}
\end{figure}

In order to determine these 11 parameters we used a standard least squares fitting method.
Using the simulated annealing approach \cite{Kirkpatrick671} it was possible to ensure that the fit corresponds to a global solution. In addition to this conventional method of analysing the data, in the Appendix~\ref{app:perturb} we develop a novel approach based on perturbation theory to facilitate the fitting procedure. Using this approach it is possible to estimate certain set of parameters  of the Hamiltonian (specifically the orientation of the $\mathbf{Q}$ tensor and the $C_2$ symmetry  axis) before performing a fitting. This in turn simplifies the search of a global solution by reducing the amount of numerical efforts for fitting procedure.

In \figref{fig:spirals} we show the experimental SHB spectra obtained using the spiral scan of the magnetic field. These maps were constructed by assembling SHB spectra as shown in \figref{fig:setup}(d) into an image, where each vertical line consists of a SHB spectrum. Each spectrum was then examined individually, in order to identify the positions of the side holes and the main, strong anti-holes. These directly give the nuclear Zeeman splittings of the excited and ground states (see \figref{fig:classes})), respectively. All measured positions can be found in \figref{fig:spirals_data} in the Appendix. The measured positions were used to fit all the parameters of the spin Hamiltonian. The final solution, which will be detailed below, accurately predicts the positions of the side holes and the main anti-holes, as shown in \figref{fig:spirals} and \figref{fig:spirals_data} in the Appendix. We further note that also the fainter anti-holes seen in \figref{fig:spirals}, which were not used for fitting, can be explained using the predicted Zeeman splittings. These anti-holes are positioned at the sum and differences of the ground and excited state Zeeman splittings (cf. \figref{fig:classes}).

It should be noted that the measured ground state Zeeman splittings are also in good agreement with predictions based on the spin Hamiltonian in Ref. \cite{Longdell2006}, up to a rotation of about 5 degrees around the $C_2$ symmetry axis (in the $D_1-D_2$ plane). This is within the estimated error of the position of the $D_1$ axis in the $D_1-D_2$ plane in Ref. \cite{Longdell2006}, which was stated to be 10 degrees.

The fitted $C_2$ symmetry axis is tilted by only 8 degrees from the $z$ axis (\tabref{table:results}), as expected from the orientation of the crystal with respect to the $z$ axis of the coils. Having identified the $C_2$ symmetry axis, it is possible to do measurements in the $D_1-D_2$ plane that is perpendicular to this axis. The results (\figref{fig:D1D2b}) are in good agreement with predicted spectra and contain only one set of lines (holes and antiholes) due to the fact that both subsites in this plane are magnetically equivalent. The degeneracy of the subsites confirms that the $C_2$ symmetry axis has been accurately determined.
 
\subsection{Fitting ambiguities due to spin Hamiltonian symmetries}

By fitting the recorded spectrum as a function of $\vec{B}_n$ one cannot determine the spin Hamiltonian without ambiguity, as there is no unique solution. This is due to the fact that the measured spectrum is invariant under certain transformations of the Hamiltonian coming from its symmetries. Some type of the symmetries related, for example, to the global rotations of the interaction tensors $\mathbf{M}$ and $\mathbf{Q}$ or the order of their diagonal elements is not physically meaningful. However, the type of the symmetry related to the relative signs of the diagonal elements (this transformation can be considered as a mirror reflection) does modify the relative orientations of the interaction tensors (for details see Appendix~\ref{app:symm}).

In general, only absolute values of the diagonal elements of the effective $\mathbf{Q}$ and $\mathbf{M}$ tensors can be extracted from the fit, which leads to the fact that relative signs of the eigenvalues can not be experimentally determined based on only such a measurement (Appendix~\ref{app:symm}). For example, for each combination of the signs of $g_1$, $g_2$ and $g_3$, one obtains different solutions that lead to the same spectrum, but for which the orientation of the $\mathbf{Q}$ tensor is different (see Appendix~\ref{app:symm}). Since the signs of the $\mathbf{M}$ tensor for the ground and excited states have never been measured for this material we have $2^3=8$ possible combinations for each state, which means a total of 64 possible solutions. 

We determined the sign of $D$ in the $\mathbf{Q}$ tensor from the known order of the zero field splittings  for $^{151}$Eu$^{3+}$:Y$_2$SiO$_5$~\cite{Lauritzen2012,Timoney2012,Jobez2015} both for the ground and excited states.

\setlength{\tabcolsep}{0.8em} 
{\renewcommand{\arraystretch}{1.5}
\begin{table}
\centering
\caption{Best fit parameters with fit errors for $^{151}$Eu$^{3+}$:Y$_2$SiO$_5$. $D$ and $E$ parameters of the quadrupole $\mathbf{Q}$ tensor were taken from previous spectroscopic studies (Ref.~\cite{Longdell2006} for ground $^7$F$_0$ and Ref.~\cite{Yano1991} excited $^5$D$_0$ states). $\gamma$ accounts for the position of the polarization extinction axes $D_1$ and $D_2$ and was measured separately using polarization dependent absorption of the crystal. All other parameters were used to fit the spin Hamiltonian on the optical transition. The error estimation was done using the covariance matrices from the nonlinear fit and  do not include errors in the magnetic field, which are expected to be less than 5\%. The angles $\alpha_i$, $\beta_i$ and $\gamma_i$ are Euler angles that express the tensors in the (X,Y,Z) lab frame.}
\label{table:results} 
\begin{tabular}{c|c|c}
parameter      & ground state, $^7$F$_0$ & excited state, $^5$D$_0$  \\ \hline\hline
$D$, MHz            & -12.3797     & 27.26           \\
$E$, MHz            & -2.735       & 5.85          \\
$\alpha_Q$, $^{\circ}$     & -29.9(3)      &165.30(7)     \\         
$\beta_Q$, $^{\circ}$      & 53.4(25)    & 154.91(35)     \\
$\gamma_Q$, $^{\circ}$     & 124.05(86)   & 107.81(45)      \\ \hline
$g_1$, MHz/T         & 4.30(12)     & 9.11(46)     \\
$g_2$, MHz/T          & 5.559(55)    & 9.158(17)     \\
$g_3$, MHz/T           & -10.891(59)   & 9.069(26)      \\
$\alpha_M$, $^{\circ}$      & 105.25(72)   & 70.53(38)    \\
$\beta_M$, $^{\circ}$       & 163.74(61)   & 5.0(2)          \\
$\gamma_M$, $^{\circ}$      & 124.56(65)   &  62.17(64)     \\ \hline
$\alpha_{C_2}$, $^{\circ}$  & \multicolumn{2}{c}{-140(4)}   \\
$\beta_{C_2}$, $^{\circ}$   & \multicolumn{2}{c}{172(3)}    \\
$\gamma$, $^{\circ}$  & \multicolumn{2}{c}{-51}     
\end{tabular}
\end{table} 
\setlength{\tabcolsep}{0.8em} 
{\renewcommand{\arraystretch}{1.5}
\begin{table}[]
\centering
\caption{\label{table:br} Comparison between predicted (cal) and measured (exp) relative optical oscillator strengths for \eu{}. The calculated values are derived from \tabref{table:results} and are compared with results from \cite{Lauritzen2012}. Rows correspond to transitions starting from the ground state hyperfine levels and columns correspond to transitions to different excited state hyperfine levels.}
\begin{tabular}{l||llll}
                            & $\ket{\pm 1/2}_e$ & $\ket{\pm 3/2}_e$ & $\ket{\pm 5/2}_e$ &        \\ \hline \hline 
\multirow{2}{*}{$\bra{\pm 1/2}_g$} & 0.02        & 0.18       & 0.80      & (calc) \\
                            & 0.03(3)    & 0.22(3)    & 0.75(3)    & (exp)  \\
                            &            &            &            &        \\
\multirow{2}{*}{$\bra{\pm 3/2}_g$} & 0.12        & 0.71     & 0.17      & (calc) \\
                            & 0.12(3)    & 0.68(3)    & 0.20(3)    & (exp)  \\
                            &            &            &            &        \\
\multirow{2}{*}{$\bra{\pm 5/2}_g$} & 0.87        & 0.10     & 0.03      & (calc) \\
                            & 0.85(3)    & 0.10(3)    & 0.05(3)    & (exp) 
\end{tabular}
\end{table}

\begin{table*}
\centering
\caption{Summary for hyperfine properties on optical \transition{} transition of \eu{} crystal for different isotopes ($^{151}$Eu and $^{153}$Eu). The $D$ and $E$ are parameters of the quadrupolar tensor $\mathbf{Q}$, $\eta=3E/D$ is the ellipticity parameter of the $\mathbf{Q}$ tensor. The nuclear magnetic moment quenching for principal values of $\mathbf{M}$ tensor is expressed using $\alpha$ parameters and the  $g_i = (1-\alpha_i)g_N$ expression, where $g_N$ is the nuclear magnetic moment of the free $^{151}$Eu$^{3+}$ ion. Experimental values for $^{153}$Eu are taken from~\cite{Yano1991,Yano1992a}.}
\label{table:isotopes} 
\begin{tabular}{c|ccccccccc}
           &           & $\nu_1$, MHz&  $\nu_2$, MHz & $D$, MHz &  $\eta$  & $\alpha_1$ & $\alpha_2$ & $\alpha_3$ \\ \hline
$^{151}$Eu & $^7$F$_0$ & 34.54 & 46.25 & -12.3797 &  0.663  &  0.59   &  0.47   &   2.03   \\
           & $^5$D$_0$ & 102  & 75   & 27.26    &  0.644     &  0.14   &   0.13   &  0.14  \\
$^{153}$Eu & $^7$F$_0$ & 90   & 119.2  & -32.02     & 0.674            &      &      &  \\
           & $^5$D$_0$ & 260  & 194  & 69.67     & 0.660           &       &      & \\
\end{tabular}
\end{table*}

\subsection{Reducing fitting ambiguities}

Some assumptions can be made to choose the global sign of both $\mathbf{M}$ tensors. The nuclear magnetic moment of the ion can be substantially quenched or even inverted due to higher order hyperfine interaction~\cite{Shelby1981}. The $\mathbf{M}^{(g)}$ tensor for the ground state is very anisotropic (see \tabref{table:results}), so its eigenvalues might differ substantially from the value of the nuclear magnetic moment of the free ion, in particular some $g$ values could even be negative. 
For the excited state $^5$D$_0$, however, this effect is negligible  \cite{Shelby1981}. This is due to the much larger energy spacing for the closest energy level for excited state ($>1700$ cm$^{-1}$ for $^5$D$_1$ while only $>200$ cm$^{-1}$ for~$^7$F$_1$) which reduces the higher order perturbation effects on the nuclear magnetic moment for~$^5$D$_0$. 

The weak perturbation in the excited state is supported by the fact that the eigenvalues of $\mathbf{M}^{(e)}$ are all similar (isotropic, see \tabref{table:results}), and close to the magnetic moment of a free ion (1.389$\mu_N=10.56$~MHz/T) up to the small quenching. Taking this into account we therefore assume that all eigenvalues of $\mathbf{M}^{(e)}$ are positive. We are then left with 16 possible solutions. 

\subsection{Identifying a unique solution from the optical branching ratios}

To find a unique solution, one could measure the quadratic Zeeman interaction using SHB, as it is sensitive to the sign of the $\mathbf{M}$ tensor~\cite{MacFarlane1981,Silversmith1986}. This approach requires measuring the shift of the spectral hole under strong magnetic fields ($\approx$1~Tesla). One can also utilize optical branching ratios which are known to be sensitive to the  sign and/or absolute value change of the nuclear projection between two electronic states. We use the latter to identify the proper solution.

The optical branching ratios at zero magnetic field were measured in a previous study using tailoring techniques~\cite{Lauritzen2012} and are given in \tabref{table:br}. We verified that the measured table of relative oscillator strengths is equivalent for the $^{151}$Eu isotope at least within the experimental errors given in Ref.~\cite{Lauritzen2012}. In order to calculate the relative oscillator strength for each transition $\ket{\pm k/2}^{(g)}\longleftrightarrow \ket{\pm l/2}^{(e)}$, we write it as an overlap between nuclear eigenstates $\mu_{eg} = \mu_{\textrm{opt}} \braket{\pm k/2^{(g)}}{\pm l/2^{(e)}}$. In this expression, $\mu_{\textrm{opt}}$ is the dipole moment of the optical transition defined by the electronic wavefunctions and is the same for each nuclear spin projection. This is done assuming  that the electronic  and the nuclear wavefunctions are separable for the ground and excited states, which was confirmed to be a good approximation in the case of quenched electronic spin  \cite{Mitsunaga1985,Bartholomew2016}. 

The branching ratio table is calculated for each magnetic subsite and the average values are considered (\tabref{table:br}). By comparing experimental  results with all possible combinations (deduced from the  assumptions discussed above) obtained from the fitted Hamiltonians we found the solution which gives the best agreement. We note that among the remaining combinations the solution given in \tabref{table:results} is the only one which gives the relative oscillator strengths close to the experimental error bars. Other possible solutions are listed in Appendix~\ref{app:symm}.

\subsection{Systematic errors and tensor orientations along $D_1$,$D_2$, and $b$ axes}

The error in the orientation of the cut surfaces of the crystal is inferior to $1\degree$. The relative orientations of the X,Y and Z axes of the coils should also be smaller than $1\degree$. The main source of error is then the orientation of the crystal with respect to the X,Y and Z axes. As discussed in Sec. \ref{fit_section}, the $C_2$ symmetry axis (or crystal $b$ axis) could be determined from the fit and it is misaligned with about $8\degree$ with respect to the Z axis. For optics experiment, the most commonly used reference frame is given by the $D_1$,$D_2$, and $b$ axes. To determine the orientation of the $D_1$ axis in the $X$-$Y$ plane we used the polarization-dependent absorption coefficient \cite{Koenz2003}. This allowed us to express the $\mathbf{M}$ and $\mathbf{Q}$ tensors in the $D_1$,$D_2$, and $b$ reference frame, which are given in \ref{app:tensors}. We estimate that the final error in the $D_1$,$D_2$, and $b$ reference frame is at most a few degrees, and mostly in the $D_1$-$D_2$ plane.

\section{Discussions and conclusions}
\label{sec:conclusion}

Our analysis yields a spin Hamiltonian which inverts the sign for one of the eigenvalues of the $\mathbf{M}^{(g)}$ tensor (see \tabref{table:br}). 
Such a sign change for the nuclear magnetic moment has been observed previously \cite{Silversmith1986}, and originates from the well established effect of nuclear magnetic moment quenching \cite{Elliott1957}. This effect is caused by the interaction with nearby $J=1$ electronic levels giving rise to the pseudoquadrupole interaction and reduced magnetic moment which can be written as $g = (1-\alpha)g_N$, where $g_N=10.56$~MHz/T is the nuclear magnetic moment of the free europium ion \cite{MacfarlaneShelby1987}. The calculated $\alpha$ parameters are given in \tabref{table:br}, both for the ground and excited states.

Another particular feature is the negative sign of the $D$ parameter in the ground state, which leads to the inverted order of energy levels at zero field (see \figref{fig:levels}(a)), while for the excited state the $D$ parameters is positive. This holds true also for the $^{153}$Eu isotope (see \tabref{table:br}). This difference in sign of $D$ has been observed in previous studies of Eu$^{3+}$ doped crystals \cite{Silversmith1986,Ahlefeldt2013b}. It can be explained by taking into account the electric field gradient created by the 4f electronic configuration in each electronic state \cite{Sharma1985,Erickson1986}. This type of contribution for Eu$^{3+}$ ion is defined by the mixing  with the second electronic level ($J=2$) but not $J=1$ due to the fact that $J=0$. This effect will be negligible for the excited $^5$D$_0$ state again due to the much higher energy  for $^5$D$_2$ levels ($>4100$~cm$^{-1}$ for $^5$D$_2$ \cite{Yano1992a} and $>860$~cm$^{-1}$ for $^7$F$_2$ levels \cite{Koenz2003}).

In conclusion, we have characterized the spin Hamiltonian of the excited state of \euiso{}. 
 We have determined all relevant parameters of the nuclear spin Hamiltonian in the  electronic  excited-state $^5$D$_0$ and characterized its orientation with respect to the ground state spin Hamiltonian. This is particularly important to be able to predict the behavior of optical transitions under external magnetic fields.  

Our characterization of \euiso{} is in good agreement with previously obtained results for relative optical strengths at zero magnetic fields. We characterized the relative signs between the hyperfine parameters for electronic ground and excited states and identified unique solution compatible with previous results  from other crystals.
Our results allow the calculation of transition frequencies and relative oscillator strengths for arbitrary magnetic field vectors. This is a crucial requirement in order to use highly coherent spin transitions in this material for the implementation of long lived optical quantum memories combined with extended spin coherence properties for spin transitions.

\section*{ACKNOWLEDGEMENTS}
The authors thank Nuala Timoney and Cyril Laplane for useful discussions, as well as Claudio Barreiro for technical support.

\section*{Funding Information}
We acknowledge funding from the Swiss FNS NCCR programme Quantum Science Technology (QSIT) and FNS Research Project No 172590, EUs H2020 programme under the Marie Sk\l{}odowska-Curie project QCALL (GA 675662) and EU’s FP7 programme under the ERC AdG project MEC (GA 339198).  \\

\bibliography{europium.bbl}

\newpage
\clearpage

\input{europium_appendix.tex}

\end{document}

%% file: europium_appendix.tex
\begin{appendix}


\section{The Q and M tensors}
\label{app:tensors}
The tensors $\mathbf{Q}_i^{(e)}$ and $\mathbf{M}_i^{(e)}$ can be diagonalized in their respective principle axis systems. In order to express them in the (X,Y,Z) lab frame we apply a rotation with the usual Euler angle convention:
\begin{equation}
\mathbf{Q}_1^{(e)}=R(\alpha_Q,\beta_Q,\gamma_Q)\cdot\begin{bmatrix}
    -E & 0 & 0 \\
    0 & E & 0 \\
    0 & 0 & D
\end{bmatrix}\cdot R(\alpha_Q,\beta_Q,\gamma_Q)^{T}  
\end{equation}
\begin{equation}
\mathbf{M}_1^{(e)}=R(\alpha_M,\beta_M,\gamma_M)\cdot \begin{bmatrix}
    g_1 & 0 & 0 \\
    0 & g_2 & 0 \\
    0 & 0 & g_3
\end{bmatrix}\cdot R(\alpha_M,\beta_M,\gamma_M)^{T},
\end{equation}
where $R(\alpha,\beta,\gamma)$ is the rotation matrix with Euler angles $(\alpha,\beta,\gamma)$ for ZYZ convention
\begin{equation} 
R(\alpha,\beta,\gamma) = R_z(\gamma)\cdot R_y(\beta)\cdot R_z(\alpha), \tag{A2bis}\label{eq:A2bis}
\end{equation}
where
\begin{align}
R_z(\alpha) = \begin{bmatrix}
    \cos (\alpha) & \sin (\alpha) & 0 \\
    -\sin(\alpha) & \cos (\alpha) & 0 \\
    0 & 0 & 1
\end{bmatrix},
\end{align}
\begin{equation}
R_y(\beta) = \begin{bmatrix}
    \cos (\beta) & 0 & -\sin (\beta)  \\
    0 & 1 & 0 \\
     \sin (\beta) & 0 & \cos (\beta) 
\end{bmatrix}.
\end{equation}

\begin{figure}
    \centering
    {\includegraphics[width=1\linewidth,trim=0.5cm 0.5cm 0.5cm 2.7cm,clip]{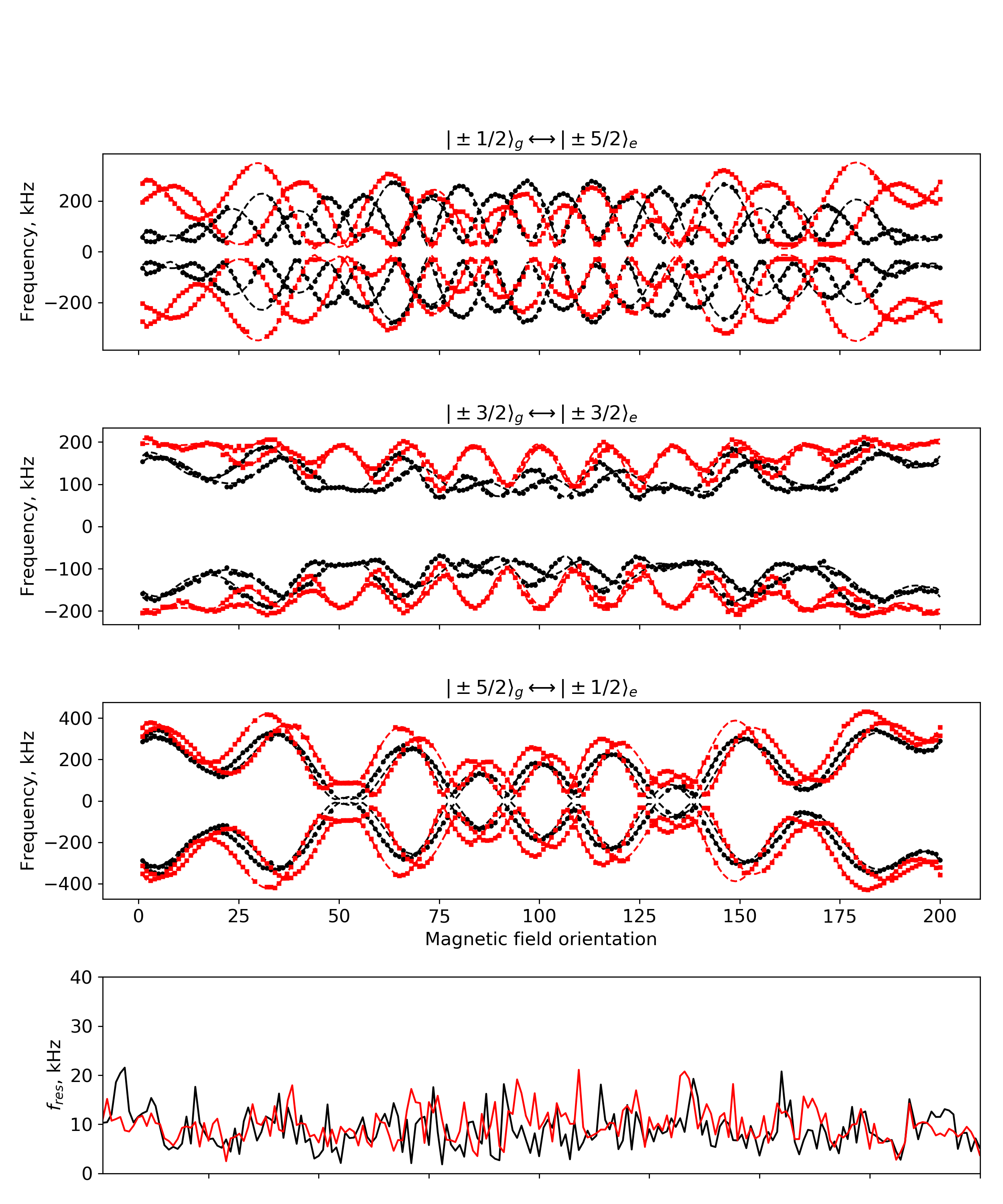}} 
    \caption{\label{fig:spirals_data}(color online). Measured positions of the side holes (red squares) and main antiholes (black circles), which correspond to the nuclear Zeeman splittings of the excited $^5$D$_0$ and ground $^7$F$_0$ state, respectively. The dashed lines represent the result of the fitting of the spin Hamiltonian for each state. The extracted parameters are presented in \tabref{table:results}.
     The bottom figure shows the average residual difference between the measured spectra and the fit (averaging over the three spectra).}
\end{figure}

The interaction tensors for the other magnetic subsite  are defined using an additional $\pi$-rotation around the symmetry axis of the crystal and given by
\begin{align}
\mathbf{M}_2^{(e)}=R_{C_2}\cdot \mathbf{M}_1^{(e)}\cdot R_{C_2}^T,\\ 
\mathbf{Q}_2^{(e)}=R_{C_2}\cdot \mathbf{Q}_1^{(e)}\cdot R_{C_2}^T,
\end{align}
where $R_{C_2}$ is the rotation of angle $\pi$ around the $C_2$ axis:
$$R_{C_2}=R^T(\alpha_{C_2},\beta_{C_2}, 0)R_z(\pi)R(\alpha_{C_2},\beta_{C_2}, 0).$$

The total rotation from the crystal $(D_1,D_2,b)$ frame to the $(X,Y,Z)$ lab frame is given by the rotation $R(\alpha_{C_2},\beta_{C_2}, \gamma)$, where $\gamma$ is an additional rotation angle in the $D_1-D_2$ plane. It was measured separately using polarization dependent absorption of the crystal.

In this work, we extract the parameters $g_1$, $g_2$, $g_3$, $\alpha_M$, $\beta_M$, $\gamma_M$, $\alpha_Q$, $\beta_Q$, and $\gamma_Q$ from the measurement of the excited state splittings, in the presence of an external magnetic field. This is done using the fitting procedure explained in the main text. The final fitting is depicted on \figref{fig:spirals_data}. The resulting  interaction tensors in the $(D_1,D_2,b)$ basis can be calculated based on the fitted parameters (\tabref{table:results}) and are found to be

$$\mathbf{Q}_1^{(g)} = \begin{pmatrix}
-3.0685  & -2.4714 &   6.7354\\ 
-2.4714 &  -4.2007 &   2.4588\\ 
 6.7354  & 2.4588   & -5.1106 
\end{pmatrix}_{D_1D_2b},$$   

$$\mathbf{M}_1^{(g)} = \begin{pmatrix}
 3.8330 &  -0.896 &  -4.7029\\
 -0.8958  &  3.3680 &  -3.7497\\
-4.7029  &   -3.7497  & -8.2410
\end{pmatrix}_{D_1D_2b},$$

$$
\mathbf{Q}_1^{(e)} = \begin{pmatrix}
4.8095 &  -1.5956 &   13.0154\\ 
 -1.5956  &  4.3611  &  7.0101\\ 
 13.0154  &  7.0101 &  18.0894
\end{pmatrix}_{D_1D_2b},$$
     
$$
\mathbf{M}_1^{(e)} = \begin{pmatrix}
  9.1340   & -0.0248 &   0.0032\\ 
  -0.0248  &   9.1347 & -0.0092\\ 
   0.0032  & -0.0092 &    9.0713
\end{pmatrix}_{D_1D_2b}.$$

\section{On the symmetry of the Hamiltonian}
\label{app:symm}

The main problem is to determine the right orientations of the magnetic tensors of the considered Hamiltonians: quadrupole interaction tensor $\mathbf{Q}$ and nuclear Zeeman interaction tensor $\mathbf{M}$. It is the relative orientations of the considered tensors for the ground and excited state which determines the optical transition strength behavior under an external magnetic field. In the case of low symmetry of the crystal site, the orientation of the interaction tensors for different energy levels can be very different. This makes the separate study of the energetic spectra of two  states to be insufficient to fully predict optical transition strengths.

Here we show that the observed data can be fitted with different orientations of the quadruple interaction tensor $\mathbf{Q}$ if the relative signs of the eigenvalues of the Zeeman nuclear interaction tensor $\mathbf{M}$ are unknown. Since the same reasoning applies to the excited state, this leads to an ambiguity about the relative orientation of the $Q$ tensors from the ground and excited state and hence to different transition probabilities.

To this end, we define the transformation $S_{i}^{\prime}$ that changes the sign of the $i$-th eigenvalue,
\begin{equation}
S_{i}^{\prime} = R(\alpha_M, \beta_M,\gamma_M) S_{i} R(\alpha_M, \beta_M,\gamma_M)^T,
\end{equation}
where $S_{i}$ is a reflection in the plane perpendicular to the $i$-th direction (i.e., $S_i$ inverts the coordinate $i$ and leaves the orthogonal components unchanged). When applied to one side of $\mathbf{M}$, this $O(3)$ transformation maps $\mathbf{M}$ to $\mathbf{M}^{\prime}$, that is, the same tensor with eigenvalues of the same moduli but with different signs. By doing a change of coordinates via $\vec{I}^{\prime} = S_{i}^{\prime T} \vec{I}$, the Hamiltonian now reads
\begin{equation}
\label{eq:1}
\mathcal{H} = \vec{I}^{\prime T}\cdot \mathbf{Q}^{\prime} \cdot \vec{I}^{\prime} + \vec{B}^{T}\cdot \mathbf{M}^{\prime} \cdot \vec{I}^{\prime},
\end{equation}
with $\mathbf{Q}^{\prime} = S_{i}^{\prime} \mathbf{Q} S_{i}^{\prime T}$. Since $\mathbf{Q}^{\prime}$ is symmetric, the transformation does not change the energy spectrum of the Hamiltonian but rotates the eigenstates.

As a consequence, for every change of sign for the eigenvalues of the $\mathbf{M}$ tensor a different orientation of the $\mathbf{Q}$ tensor is found, which in total gives the same experimental spectra. The list of possible solutions for the ground and excited states is given in~\tabref{table:signs}. In general, eight different combinations of the signs lead to eight different solutions for each state (some of them can be equivalent). This leads to $8\times 8$ possible ways to connect each pair of solutions to calculate the branching ratio table. 

\setlength{\tabcolsep}{0.99em} 
\renewcommand{\arraystretch}{1.5}
\begin{table*}[]
\centering
\caption{\label{table:signs} The list of possible solutions for the ground state Hamiltonian for different combination of signs of the $\mathbf{M}$ tensor for the ground or excited states. Assuming positive signs for the excited state $\mathbf{M}^{(e)}$ tensor (first solution), the solution 4 for the ground state was found to be consistent with the optical branching ratio measurements.  }
\begin{tabular}{c|ccc|ccc|ccc}
\multirow{2}{*}{Solution} & \multicolumn{3}{c|}{$\mathbf{M}$ tensor signs} & \multicolumn{3}{c|}{$\mathbf{Q}^{(g)}$ tensor angles}                 & \multicolumn{3}{c}{$\mathbf{Q}^{(e)}$ tensor angles}                  \\ \cline{2-10} 
                          & $g_1$          & $g_2$          & $g_3$          & $\alpha_Q$, $^{\circ}$ & $\beta_Q$, $^{\circ}$ & $\gamma_Q$, $^{\circ}$ & $\alpha_Q$, $^{\circ}$ & $\beta_Q$, $^{\circ}$ & $\gamma_Q$, $^{\circ}$ \\ \hline
1                         & +              & +              & +              & --149.96                & 93.88                 & 124.10                 & 165.2982  &  154.9117   & 107.8092     \\
2                         & --              & +              & +              & 157.85                 & 95.76                 & 97.23                  & 191.8467     & 151.8768   & 335.2023    \\
3                         & +              & --              & +              & 140.59                 & --124.22               & 88.90                  & 212.0108     &  149.7404  &  172.8981   \\
4                         & +              & +              & --              & --29.90                 & 53.48                 & 124.05                 & 28.1173   &     32.8277    &   96.0319  \\
5                         & --              & +              & --              & 39.41                  & 55.78                 & 91.10                  & 327.9892  &   30.2596  &  352.8981   \\
6                         & +              & --              & --              & 22.14                  & 84.24                 & --82.77                 & 348.2384   &   28.0900  &  155.1711  \\
7                         & --              & --              & +              & --150.10                & 126.52                & --55.95                 & 151.8827  &  147.1723  &  276.0319  \\
8                         & --              & --             & --              & --30.04                 & 86.12                 & --55.90                 & 11.5272 &    25.0883 &  287.8092                               
\end{tabular}
\end{table*}

\section{Perturbation theory approach}
\label{app:perturb}

While the search for the parameters of $\mathcal{H}_{\mathrm{}}$ (Eq.~\eqref{eq:Heff}) as presented in \tabref{table:results} was done numerically for the exact Hamiltonian, a perturbation theory approach helps to better understand the energy splittings as a function of the  magnetic field orientation. It  can also be used to facilitate the fitting procedure of the measured spectra involving 11 parameters for another material. The perturbation approach can be used to estimate certain number of parameters which can be further used as an initial guess for the nonlinear fitting. This can substantially decrease the computational time and verify its consistency. In our case, the quadrupole interaction 
$\mathcal{H}_0 = \vec{I}^{T}\cdot \mathbf{Q} \cdot \vec{I}$
is dominant over the Zeeman term $\mathcal{H}_{1} = \vec{B}^{T}\cdot \mathbf{M} \cdot \vec{I}$, Eq.~\eqref{eq:Heff} (again we only consider one subsite). Then, at the first order, the energy splitting of each degenerate level $k \in \left\{ 1/2,3/2,5/2 \right\}$ can be seen as an isolated two-level system. Let us denote the eigenstates of $\mathcal{H}_0$ by $| \pm k \rangle$. The energy splitting is approximately 
\begin{equation}
\delta _k \approx | \left\langle + k \right| \mathcal{H}_1 \left| +k \right\rangle  -  \left\langle - k \right| \mathcal{H}_1 \left| -k \right\rangle | .
\end{equation}
Note that the degeneracy in $\mathcal{H}_0$ leads to an apparent ambiguity in the choice of $| \pm k \rangle$. For a well defined perturbation theory calculation, we have to ensure that $\left\langle -k \right| \mathcal{H}_1 \left| +k \right\rangle = 0$ implying that we have to maximize $\delta _k$ over all possible eigenbases for each subspace $k$. In other words, one finds that $\delta _k = \lambda_{k}^{+} - \lambda_{k}^{-} = 2 \lambda_{k}^{+}$, where $\lambda_{k}^{+}$ and $\lambda_{k}^{-}$ are the maximal and minimal eigenvalues, respectively, of $\mathcal{H}_1$ reduced to the subspace spanned by $| \pm k \rangle $.

\begin{figure*}[htbp]
  \centerline{
\includegraphics[width=0.7\columnwidth]{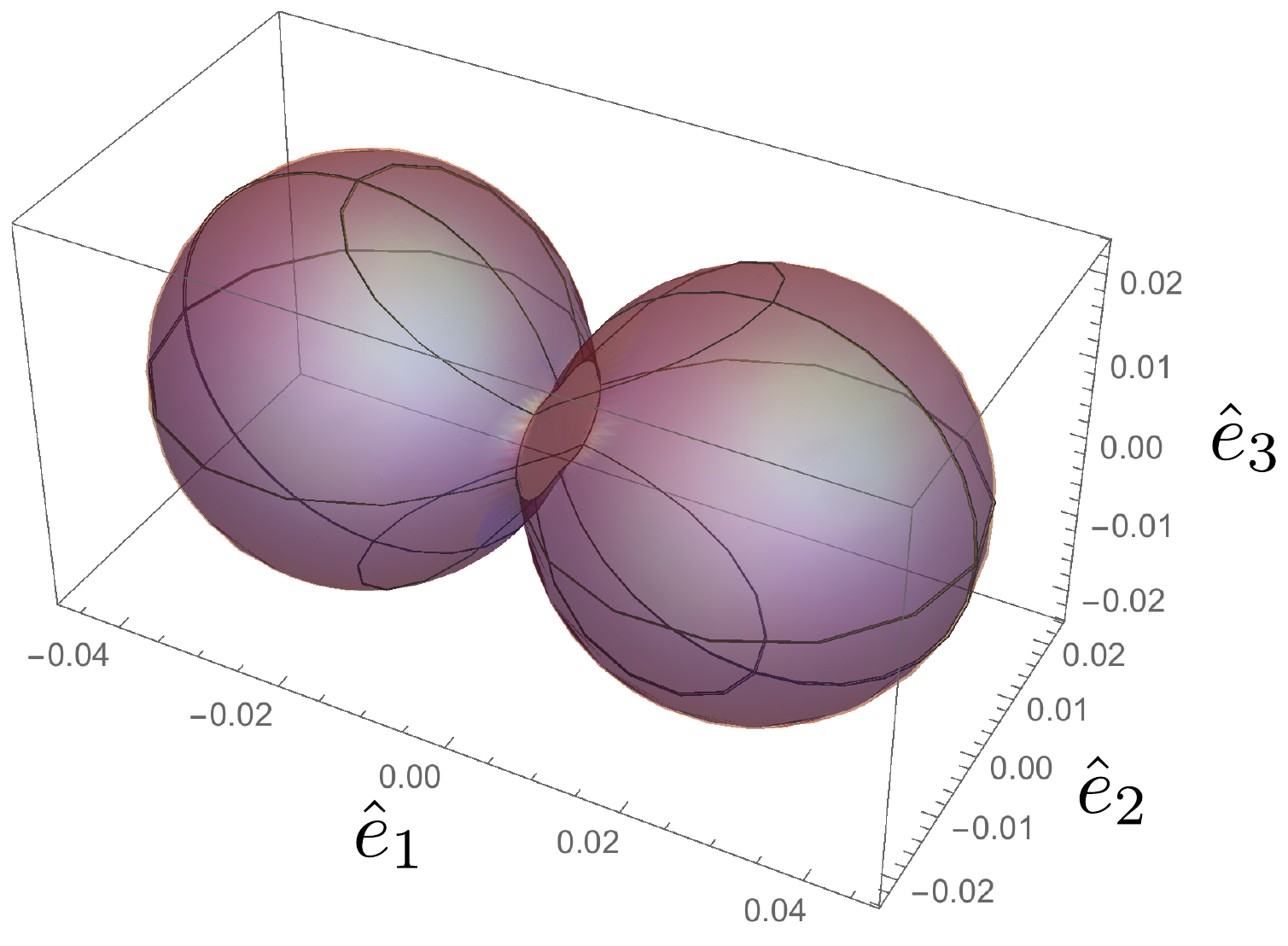}
\includegraphics[width=0.5\columnwidth]{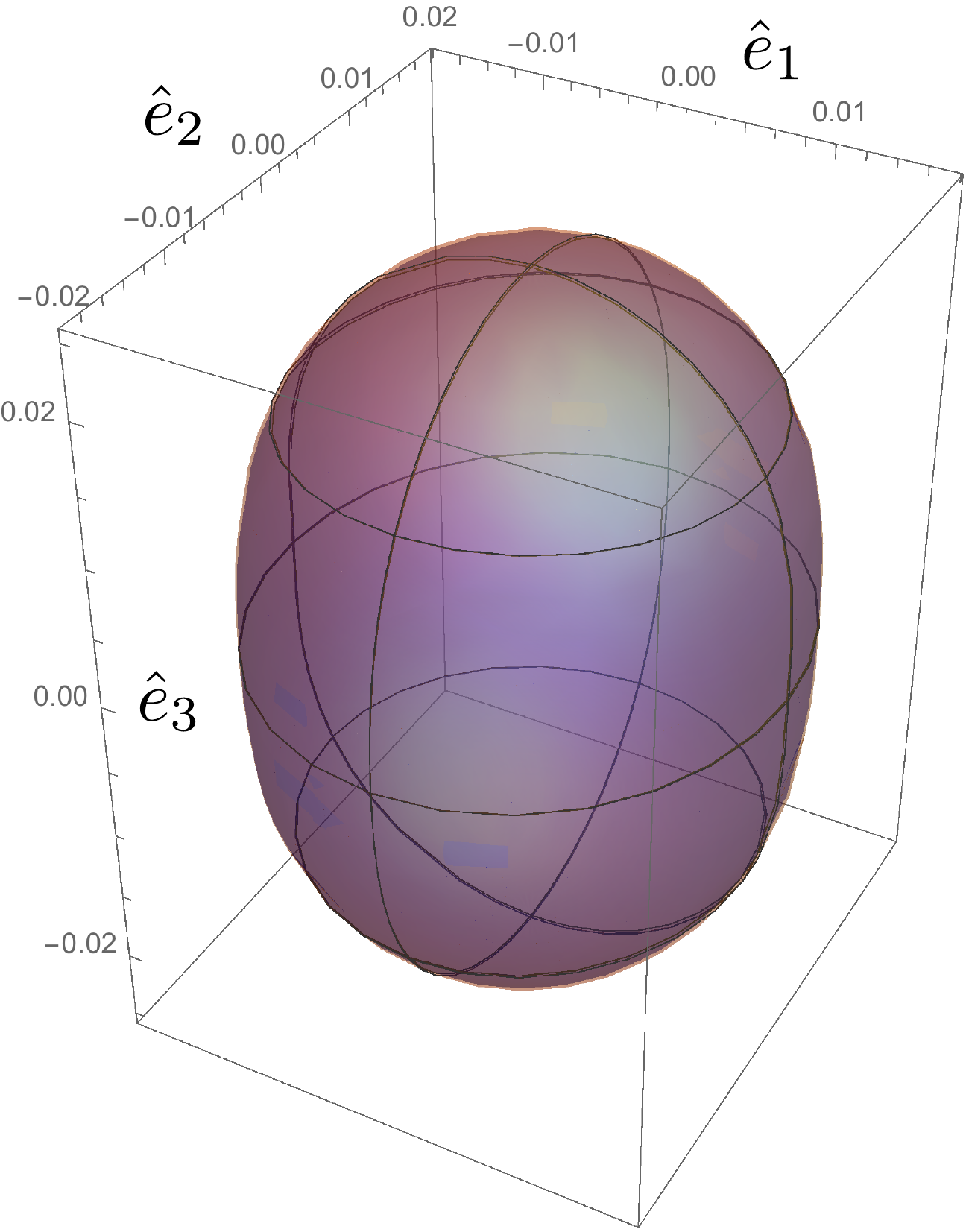}
\includegraphics[width=0.5\columnwidth]{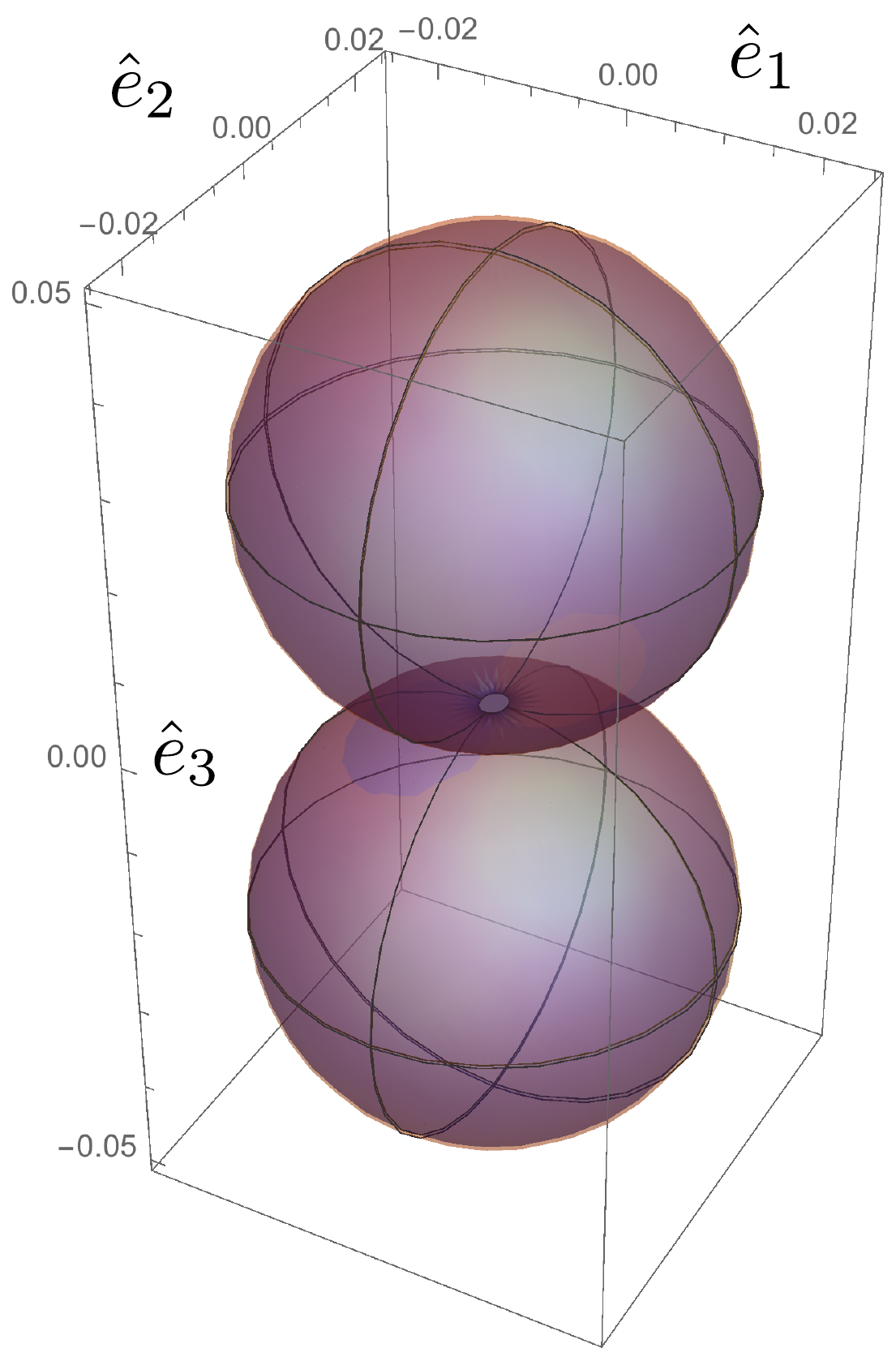}}
\includegraphics[width=0.7\columnwidth]{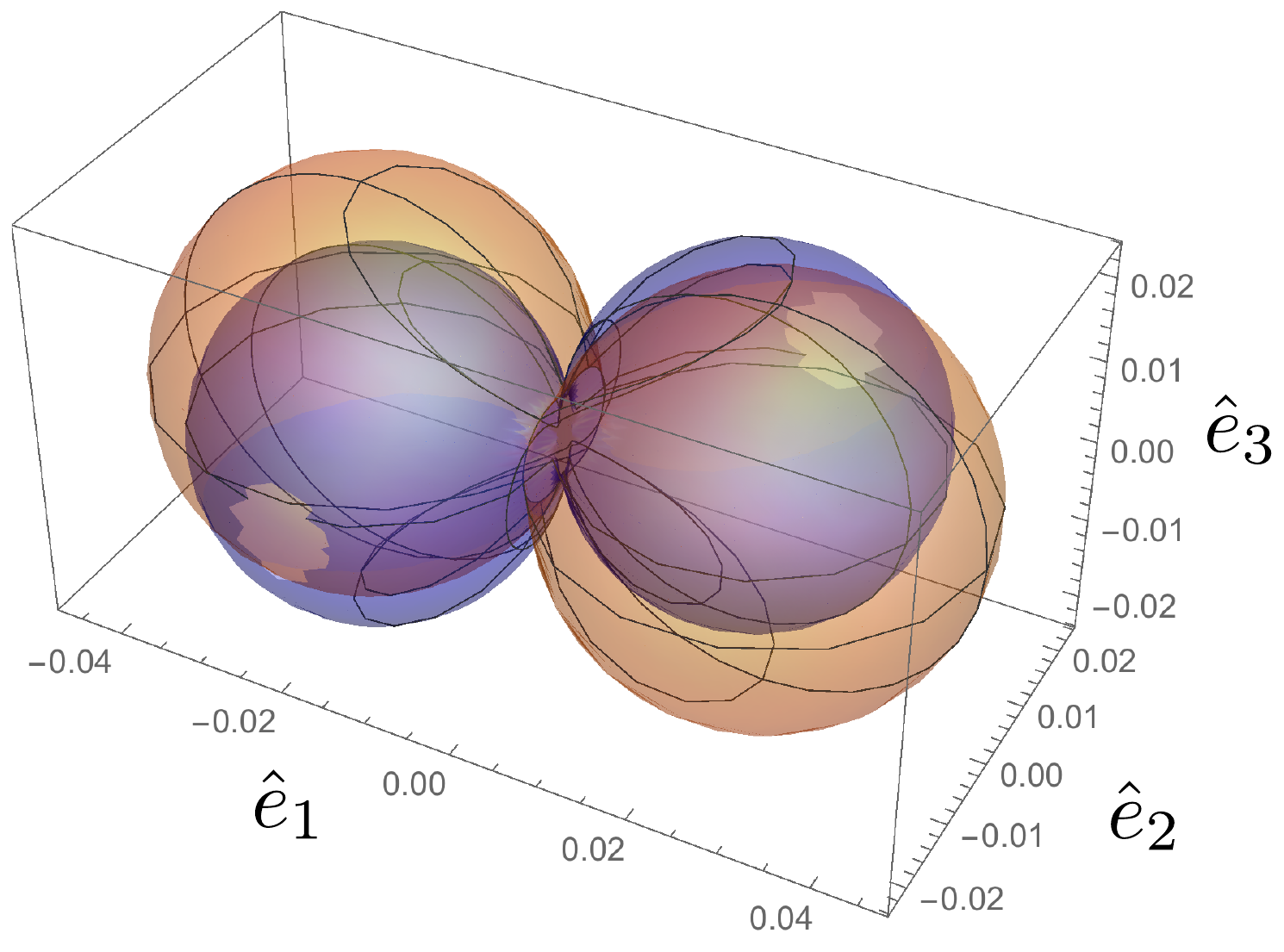}
\includegraphics[width=0.5\columnwidth]{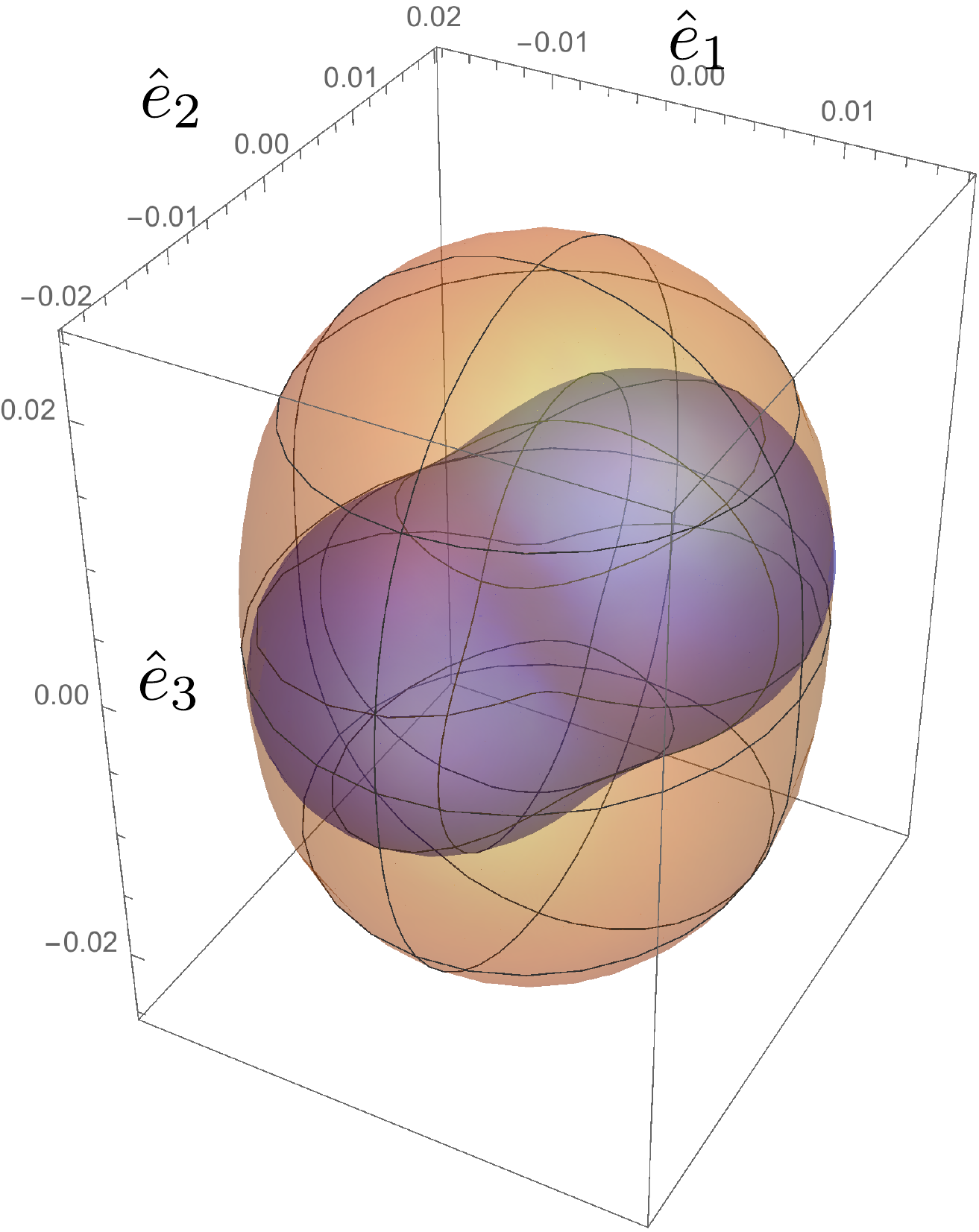}
\includegraphics[width=0.5\columnwidth]{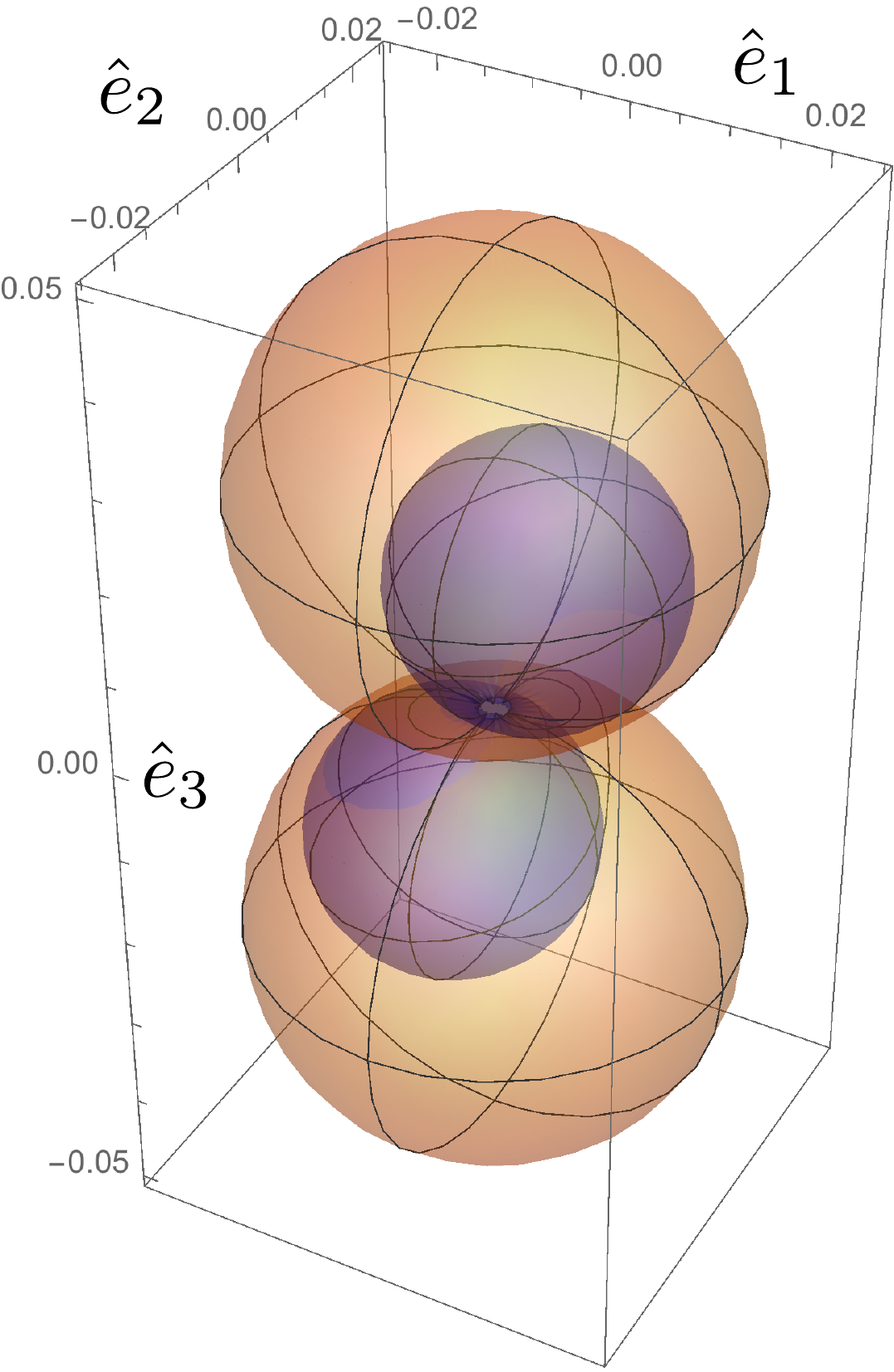}
\caption[]{\label{fig:perturbation} Spherical plot (violet) of the energy splitting for the ground (three figures below) and excited (three figures above) states, $\delta_k/(2 |\vec{B}|)$, in natural units for $k =  1/2,3/2,5/2$ (from left to right) as a function of $\vec{n}(\theta,\phi) = \vec{B}/|\vec{B}|)$. The orange plot is the hypothetical energy splitting if $M$ were isotropic (i.e., $M \propto \mathbbm{1}$), which is almost the case for excited state (and, hence, there the orange plot is basically covered by the violet one). The coordinate system is the eigenbasis of $\mathbf{Q}^{(e)}$ or $\mathbf{Q}^{(g)}$, respectively.}
\end{figure*}

Let us discuss the special case of isotropic coupling, $\mathbf{M} = \mathbbm{1}$. In addition, we work in a reference frame $\vec{I} = \sum_{i = j }^{3}I_j \hat{e}_j$ where $\mathbf{Q}$ is diagonal (i.e., $\mathbf{Q} = -E \hat{e}_1 \cdot \hat{e}_1^T + E \hat{e}_2 \cdot \hat{e}_2^T
+ D \hat{e}_3 \cdot \hat{e}_3^T$) and denote the direction of $\vec{B}$ in this frame by $\vec{n}(\theta,\phi) = \sin \theta (\cos \phi \hat{e}_1 + \sin \phi \hat{e}_2) + \cos \theta \hat{e}_3$. Then, it turns out that 
\begin{equation}
\lambda_{k}^{+}(\theta,\phi) = \sqrt{\sum_{j = 1}^3 c_{k,j}^2 n_j(\theta,\phi)^2},
\end{equation}
where the $\left\{ c_{k,j} \right\}_j$ only depends on $k$ and the eigenvalues of $\mathbf{Q}$: $\pm E,D$. In other words, the energy splitting $\delta _k$ is proportional to the distance from the origin to the surface of an ellipsoid with principal axes aligned to the eigenbases of $\mathbf{Q}$ and length $c_{k,j}$. 
Some examples, which were calculated for the extracted spin Hamiltonians, are depicted in \figref{fig:perturbation}.

\begin{figure*}[htbp]
\centerline{\includegraphics[width=.7\columnwidth]{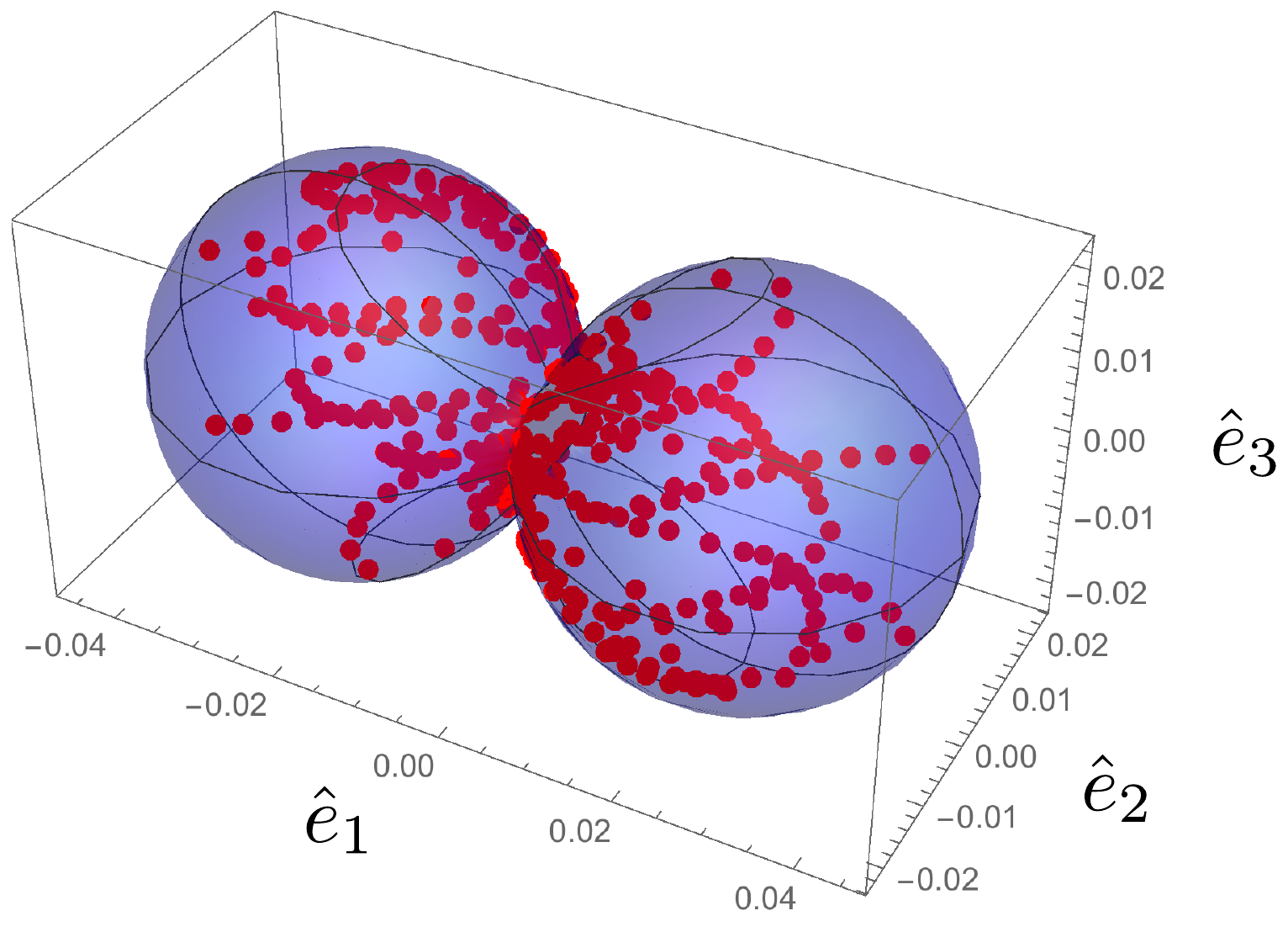}\includegraphics[width=.5\columnwidth]{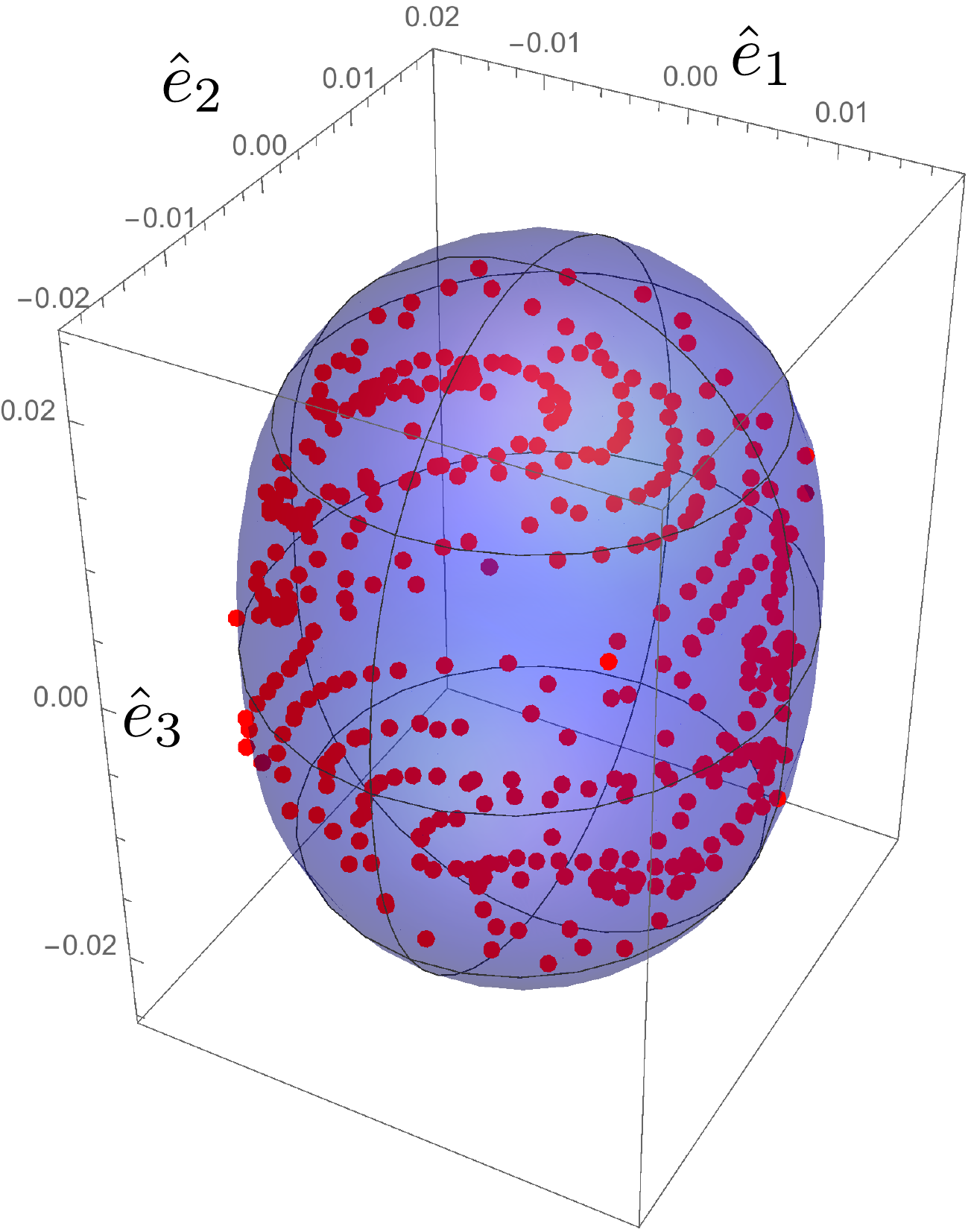}
\includegraphics[width=.5\columnwidth]{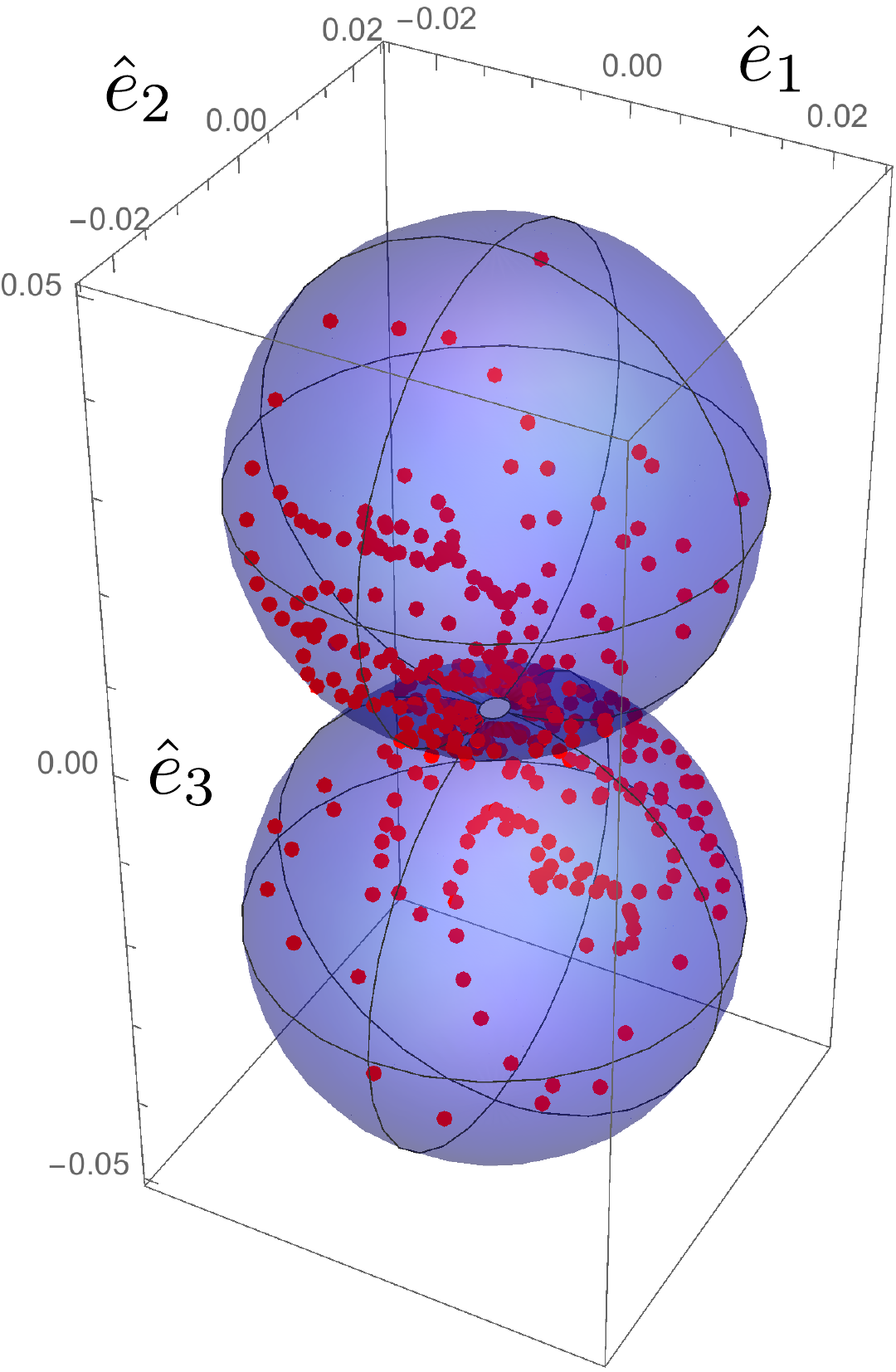}}
\includegraphics[width=.7\columnwidth]{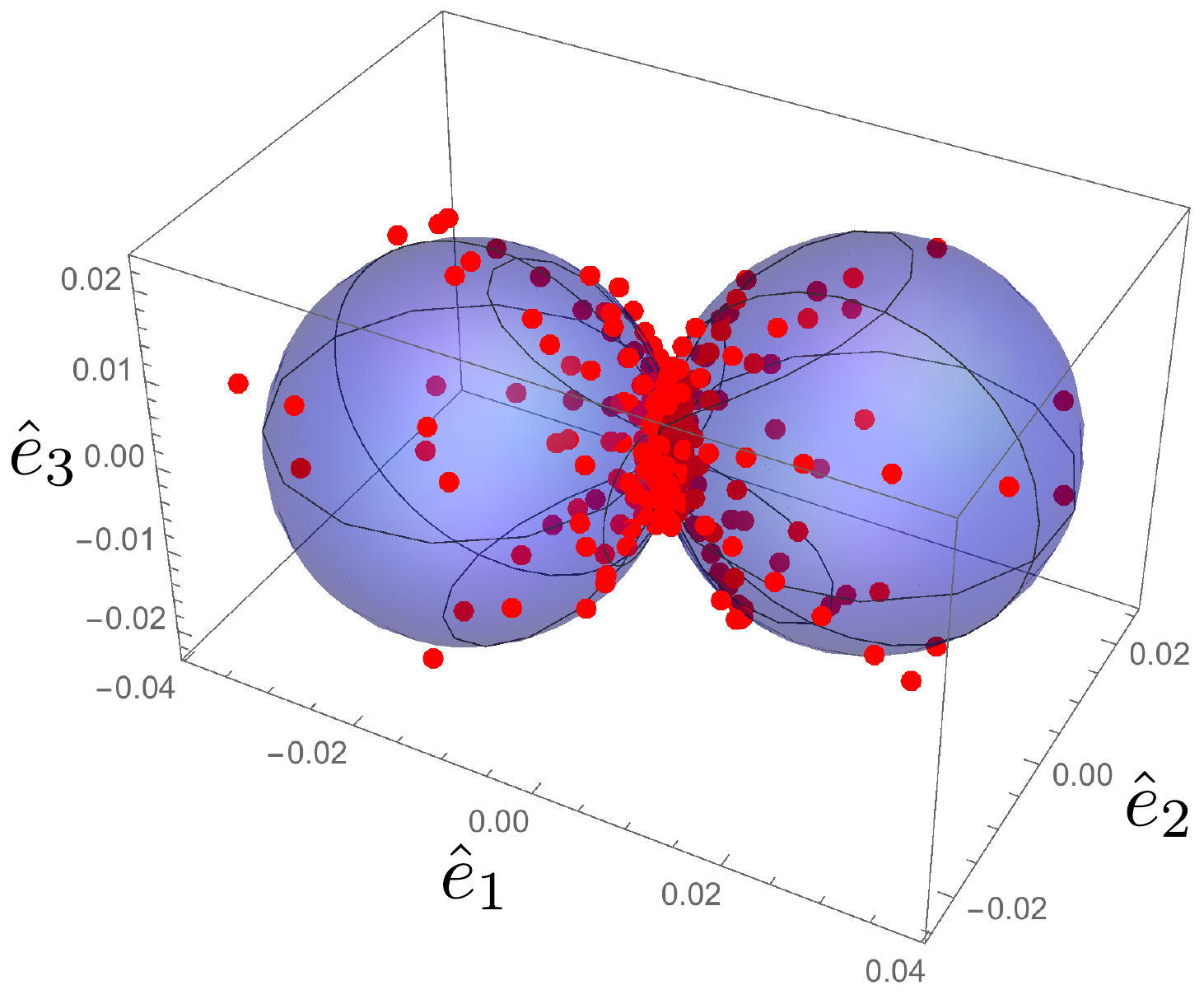}
\includegraphics[width=.5\columnwidth]{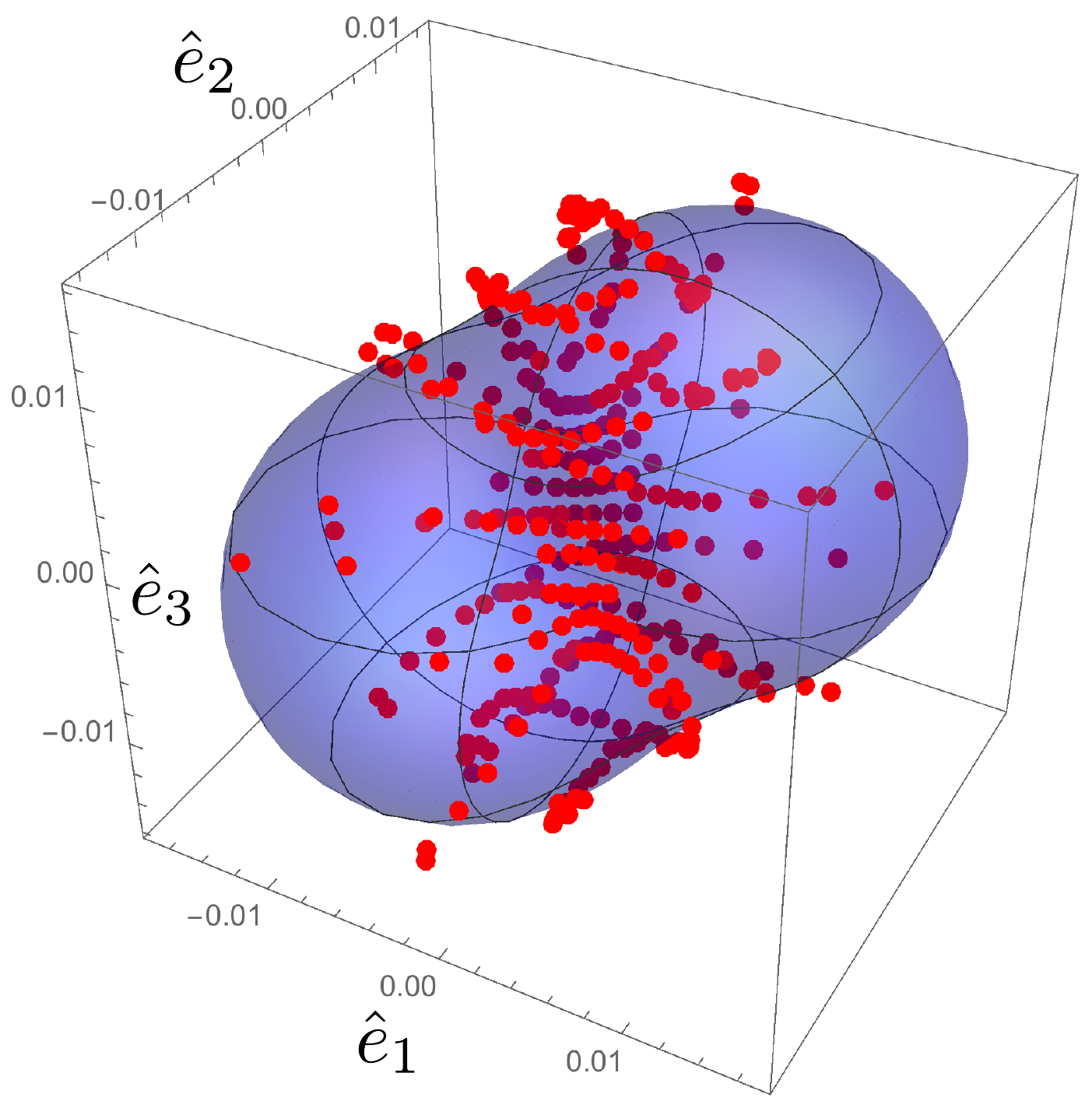}
\includegraphics[width=.5\columnwidth]{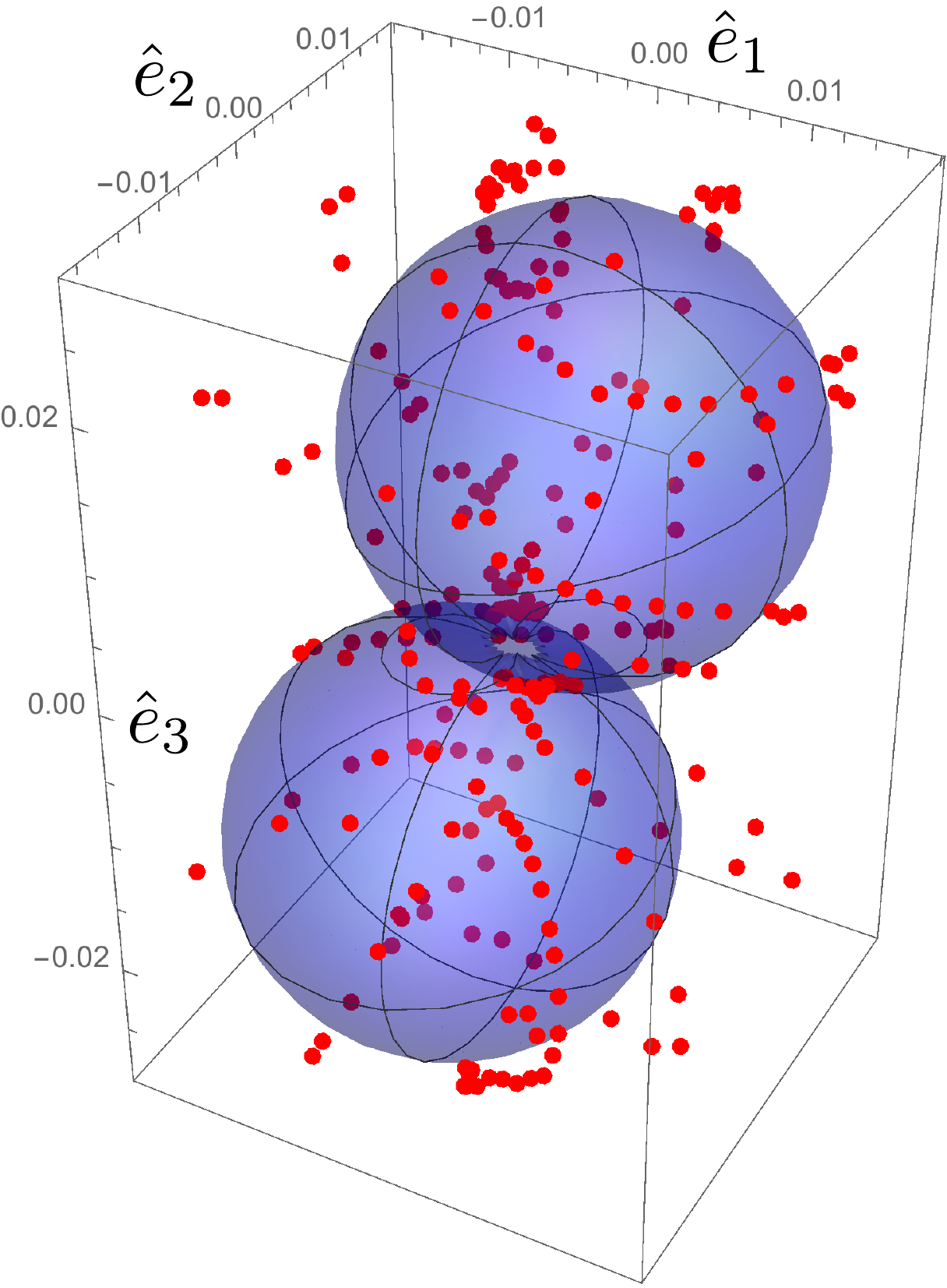}
\caption[]{\label{fig:overlayExpTheory} Overlay of $\delta _k$  for $k =  1/2,3/2,5/2$ (from left to right) from the experimental data (red dots) and the perturbation theory based on the fitted parameters (blue surface) in the laboratory frame for the ground (three figures below) and excited (three figures above) states. }
\end{figure*}

In the laboratory frame, the principal axes of the ellipsoids are rotated by $R(\alpha_Q,\beta_Q,\gamma_Q)$. If we could assume $\mathbf{M} = \mathbbm{1}$, we could directly identify the unknown angles $\alpha_Q,\beta_Q,\gamma_Q$ from the orientation of the ellipsoids in the laboratory frame. This does no longer hold in the case of general $\mathbf{M}$. In our case, it turns out that $|\mathbf{M}| \approx \mathbbm{1}$, which means that the orientation of $\mathbf{Q}^{\prime}$ (see Appendix~\ref{app:symm}) is close to the orientation given by the measured values of~$\delta _k$ (see \figref{fig:overlayExpTheory}).

In summary the procedure to fit the spin Hamiltonian of the form can be described in different steps:
\begin{enumerate}
\item The parameters $D$ and $E$ of the $\mathbf{Q}$ tensor can be determined from broadband SHB or RHS.
\item Measuring all the splittings in different directions for each energy level and using perturbation approach, one can estimate the $\mathbf{Q}$ tensor angles from the orientation of the ellipsoids in the laboratory frame as described above. From this, all the required parameters of the $\mathbf{Q}$ tensor are found.
\item For our crystal, due to the presence of two magnetic subsites, it was necessary to deduce the orientation of the symmetry axis $C_2$. This orientation can be estimated precisely by looking at the measured spectras and choosing directions of the magnetic field where two splittings coincide or are very close to each other. By extracting their positions, it is possible to get the orientation of the symmetry axis.
\item From this point, the only parameters which are unknown correspond to the $\mathbf{M}$ tensor. Six parameters representing three eigenvalues and three rotation angles can be used to fit the data assuming that all its eigenvalues are in the order of magnitude of the nuclear magneton $\mu_N$. In this way only six parameters can be used for the first fit which highly simplifies the overall task.
\end{enumerate}
We approved this procedure for our case. It substantially decreased the numerical effort in a problem with a large number of free parameters.

\end{appendix}